\documentclass[lettersize,journal]{IEEEtran}
\usepackage{amsmath,amsfonts}
\usepackage[ruled,linesnumbered]{algorithm2e}
\usepackage{array}
\usepackage{enumitem}
\usepackage{subfigure}
\usepackage{textcomp}
\usepackage{tabularx}
\usepackage{stfloats}
\usepackage{url}
\usepackage{verbatim}
\usepackage{graphicx}
\usepackage{hyperref}
\hypersetup{colorlinks=true,citecolor=black,linkcolor=black,urlcolor=black}
\usepackage[hyperref=true,backend=biber,sorting=none,backref=true]{biblatex}
\usepackage{multirow}
\usepackage{grumble}
\begin{document}

\title{DPIFrame: A Dual-Level Parallelism Acceleration Framework for CTR Model Inference}

\author{
    Dezhi Yi, 
    Huifeng Guo, 
    Kunpeng Xie, 
    Zhaolong Jian, 
    Haochi Yu, 
    Wenxuan He, \\
    Zhenhua Dong, 
    Ruiming Tang, 
    Ye Lu
    \thanks{This work was supported by the National Natural Science Foundation of China (No. 62372253) and the Fundamental Research Funds for the Central Universities (No. 079-63263260). \textit{(Corresponding author: Ye Lu.)}}
    \thanks{Dezhi Yi, Kunpeng Xie, Zhaolong Jian, Haochi Yu, and Wenxuan He are with College of Computer Science, Nankai University, Tianjin 300350, China, also with Tianjin Key Laboratory of Network and Data Science Technology, Tianjin 300350, China, and also with Key Laboratory of Data and Intelligent System Security, Ministry of Education, Tianjin 300350, China (e-mail: dezhi.yi@mail.nankai.edu.cn; xkp@mail.nankai.edu.cn; jianzhaolong@mail.nankai.edu.cn; yuhaochi@mail.nankai.edu.cn; wxhe@mail.nankai.edu.cn).}
    \thanks{Huifeng Guo and Zhenhua Dong are with Huawei Technologies Co., Ltd., Shenzhen 518129, China (e-mail: huifeng.guo@huawei.com; dongzhenhua@huawei.com).}
    \thanks{Ruiming Tang is with Kuaishou Technology Co., Ltd., Beijing 100085, China (e-mail: tangruiming@kuaishou.com). This work was done while Ruiming Tang was with Huawei Technologies Co., Ltd.}
    \thanks{Ye Lu is with College of Cryptology and Cyber Science, Nankai University, Tianjin 300350, China, also with Tianjin Key Laboratory of Network and Data Science Technology, Tianjin 300350, China, and also with Key Laboratory of Data and Intelligent System Security, Ministry of Education, Tianjin 300350, China (e-mail: luye@nankai.edu.cn).}
}

\markboth{Journal of \LaTeX\ Class Files,~Vol.~14, No.~8, August~2021}%
{Shell \MakeLowercase{\textit{et al.}}: A Sample Article Using IEEEtran.cls for IEEE Journals}


\maketitle

\begin{abstract}
Deep learning technology has enhanced the ability of Click-through rate (CTR) prediction models to learn features and improve prediction accuracy. 
However, it is challenging to deploy CTR models on GPU smoothly and perform inference efficiently, because there is a huge mismatch between the serial computational pattern and the parallel model structure. 
In this paper, we propose \ours, the first dual parallelizable framework to accelerate CTR model inference. 
In \ours, a) a dual parallelizable architecture is proposed to perform parallel CTR model inference in both intra-module and inter-module; 
b) an efficient multi-table lookup algorithm is presented for embedding operations through anticipating the whole workload in advance;
c) a breadth-first stream scheduling strategy is designed for fine-grained management of parallel computation on GPU to further supporting the dual parallel execution. 
Extensive experiments are conducted on two real-world datasets, and the results highlight that \ours~can reduce the embedding latency efficiently by \textbf{23.0$\times$} compared to PyTorch. 
Compared with PyTorch, TorchRec, HugeCTR, and OneFlow, \ours~can achieve state-of-the-art inference performance on GPU with speedups of \textbf{5.83$\times$}, \textbf{4.29$\times$}, \textbf{2.15$\times$}, and \textbf{2.0$\times$}, respectively. 
\end{abstract}


\section{INTRODUCTION}

Industrial recommender systems usually involve millions of advertisers, billions of users, and even nearly a trillion dollars in revenues~\cite{yang2022click}.\
The most important part in those recommender systems is the Click-through rate (CTR) prediction model, since it can improve the user experience effectively~\cite{kaasinen2009user} and increase advertising incomes greatly~\cite{wang2020survey}.\
Modern typical CTR model structure generally includes two key parts, the embedding module, which consists of independent sparse operations~\cite{liu2024embedding, li2024ndrec}, and the neural network module, which consists of explicit interaction layers and implicit interaction layers~\cite{yi2026eenet}.\
The frequently used operators, such as vector operations in both types of layers, can help calculate features, enhance representation features, and establish correlation relationships effectively.\
By integrating the embedding module with the neural network module, the CTR model can leverage these benefits to achieve higher inference accuracy~\cite{fu2025unified}.

However, as the number of operators and layers inside both the neural network module and embedding module increases significantly, the deep and complex structures of CTR models often result in parameter data explosion by hundreds or even thousands of times.\
To handle these large amounts of data, the acceleration hardware GPU is widely deployed~\cite{acun2021understanding, wang2024oper}, which can leverage CUDA kernels to enhance the processing of model computation tasks~\cite{baji2018evolution}.\
Nevertheless, the CTR model inference on the GPU needs to satisfy several stringent performance requirements.
The first is the stringent accuracy demand~\cite{li2023adaptive}.\
Even a 0.1\% error on the CTR model inference accuracy can lead to hundreds of millions of dollars in economic losses in advertising recommender applications ~\cite{guo2017deepfm, wang2017deep, wang2021dcn}.\ 
The second is the strict response latency limitation~\cite{pan2023recom}.\
To process the escalated data daily, recommender systems usually require fast model inference speed, which is typically in the millisecond range~\cite{wang2023bert4ctr, li2023fragment}.\
The third is the high hardware computing resource utilization requirement~\cite{pan2024recflex}.\
The operations in the aforementioned two modules are made up of massive short tasks.\
When CTR model inference, they critically require a reasonable parallel mechanism to improve GPU hardware utilization to avoid wasted investment~\cite{gupta2021recpipe}, but even though the most popular PyTorch~\cite{paszke2019pytorch} framework can only achieve under 40\% single GPU utilization, as our previous actual profiling.\
Thus, it is so challenging to take into account the three above aspects at the same time.

Existing approaches struggle to balance those three goals.
Compression techniques such as quantization~\cite{li2023adaptive, xu2021agile}, pruning~\cite{deng2021deeplight, qu2022single}, and hashing~\cite{desai2022random, pansare2022learning} 
can effectively reduce the computational demands of the CTR model.\
However, those aggressive compression methods will significantly degrade the accuracy of the model, thus making the model unusable for industrial recommendation systems~\cite{hooker2019compressed}.\
Additionally, data parallelism~\cite{zhao2020distributed, gupta2021training, zeng2024accelerating}, model parallelism~\cite{mudigere2022software, huang2025traci}, and hybrid parallelism~\cite{adnan2024heterogeneous, wang2024rap} techniques are deployed across multiple GPUs to improve the computational efficiency of the CTR model.\
Their computation scheduling strategies are typically applied at the level of computation graph, module, or layer, rather than the finer-grained operator level.\
This limitation, coupled with the communication overhead, frequently leads to insufficient GPU utilization.\
So, optimizing the efficiency of models at the algorithm level or increasing the number of devices simply cannot achieve the balance goals among the three aspects.\
To this end, we should turn the perspective on the underlying architecture for the CTR models in the recommender system and overcome the following threefold challenges to meet the aforementioned performance requirements.\

Firstly, the sparse and irregular access pattern of the embedding module conflicts with the execution model of GPU.\
Recommendation models exhibit fundamental architectural differences from traditional machine learning models, with over 95\% of their parameters concentrated in the embedding module.\
It requires independent sparse lookup operations across tens to hundreds of feature fields, which are inherently mismatched with the dense, compute-intensive execution model of GPU.\
As industrial-scale data continues to grow exponentially, these discrete and sparse embedding lookup operations have emerged as the primary performance bottleneck.

Secondly, the frequent and fine-grained vector operations in CTR models incur considerable launch overhead and fail to exploit GPU parallelism fully.
The explicit interaction module in recommendation models involves largely frequent vector operations to capture complex feature relationships.\
This computation pattern introduces substantial operator launch overhead and redundant memory accesses, thus limiting the potential for parallel execution with operators in the implicit interaction module.
For instance, in our previous profiling, the actual execution time for a single lookup operator is often about 4.75$\mu s$, but the overhead of launching the operator is about 21.92$\mu s$.\ 
So many small computational workloads are too brief and insufficient for GPU to accelerate computation by parallelism.\ 

Thirdly, the coarse-grained scheduling mechanisms of existing frameworks prevent effective operator-level parallelism in CTR models.
Both the explicit and implicit interaction modules in recommendation models exhibit a parallel characteristic from an architectural perspective, theoretically enabling parallel execution via multiple CUDA streams.\ 
However, due to the relatively low computation of individual operators in recommendation models, the existing deep learning frameworks which primarily employ coarse-grained control mechanisms, struggle to effectively parallelize those operations.\
For example, although the PyTorch framework can provide the advanced APIs to create distinct CUDA streams~\cite{choi2021implementing}, this kind of scheduling orients the stream-wise rather than the operator-wise.
It makes CUDA stream parallel execution still cannot be controlled according to higher-level logic, due to considerations of computational stability and result determinacy by PyTorch~\cite{czarnul2020investigation}. 
Besides, the scheduling strategy PyTorch exploited follows a depth-first principle by default~\cite{kwon2020nimble}, which can cause the different operators in various streams to miss earlier startup opportunities, thus resulting in the CTR model's low performance. 
A reasonable choice is the breadth-first scheduling strategy, which can alternate the launching of operators across different streams, but it is too difficult for PyTorch to precisely schedule operators. 

To overcome the above difficulties, we propose DPIFrame, which is the first dual-level parallelism acceleration framework for CTR model inference in industrial recommender systems.\
Aiming to boost inference performance, we redesign and optimize the underlying system architecture and mechanism by addressing the root causes of inefficiency in GPU-based CTR inference.\
Our contributions are summarized as
follows:\

\begin{itemize}

    \item A parallel execution architecture that operates both intra- and inter-module is proposed to close the gap between serial computational patterns and parallel model structures in CTR inference on GPU. The dual parallelizable architecture innovatively improves CTR inference performance from an underlying system-level perspective.
    
    \item An efficient and novel multi-table lookup algorithm is presented for embedding operations by anticipating their whole workload in advance. 
    Based on our designed output-first workload allocation, we can ensure address continuity during data read/write. 
    Embedding latency compared with PyTorch can be reduced greatly by an average of \textbf{12.9$\times$} and a maximum of \textbf{23.0$\times$}. 
    
    \item An operator-wise stream scheduling strategy at system level with breadth-first principle is designed to further supporting intra- and inter- module parallelism, thus deeply addressing the mismatch between hardware execution and model structure. 
    The alternately overlapping scheduling strategy help DPIFrame achieve operator-level parallel execution, the maximum GPU utilization improvement can be of \textbf{36.4\%} and \textbf{60.6\%} on Avazu and Criteo datasets, respectively.
    
    \item DPIFrame is implemented as a building block and is embedded seamlessly into PyTorch. 
          Compared to original PyTorch, TorchRec, HugeCTR, and OneFlow, \ours~can achieve end2end inference speedup by \textbf{5.83$\times$}, \textbf{4.29$\times$}, \textbf{2.15$\times$}, and \textbf{2.0$\times$}, respectively.
    
\end{itemize}

\section{BACKGROUND}
\label{sec: Background}
In this section, we first introduce the fundamental concepts related to recommender systems and CTR model.
For clarity, the typical DCN~\cite{wang2017deep} is shown in Figure~\ref{fig: fig1} as an example to illustrate the preliminaries and CTR model characteristics.
We then introduce and analyze CUDA stream mechanism.

\subsection{Recommender System}
Recommender systems have become foundational in modern digital platforms, including e-commerce, online advertising, content streaming, and social networks.
Their primary goal is to predict user preferences and provide personalized content, thereby improving user experience and increasing business revenue.
A core task in these systems is CTR prediction, which estimates the probability that a user will interact with a recommended item, such as clicking on a product or watching a video. 
Accurate CTR prediction is crucial, especially in scenarios where only a limited number of items can be recommended within strict time constraints.

\begin{figure}
\centering
  \includegraphics[width=1.0\linewidth]{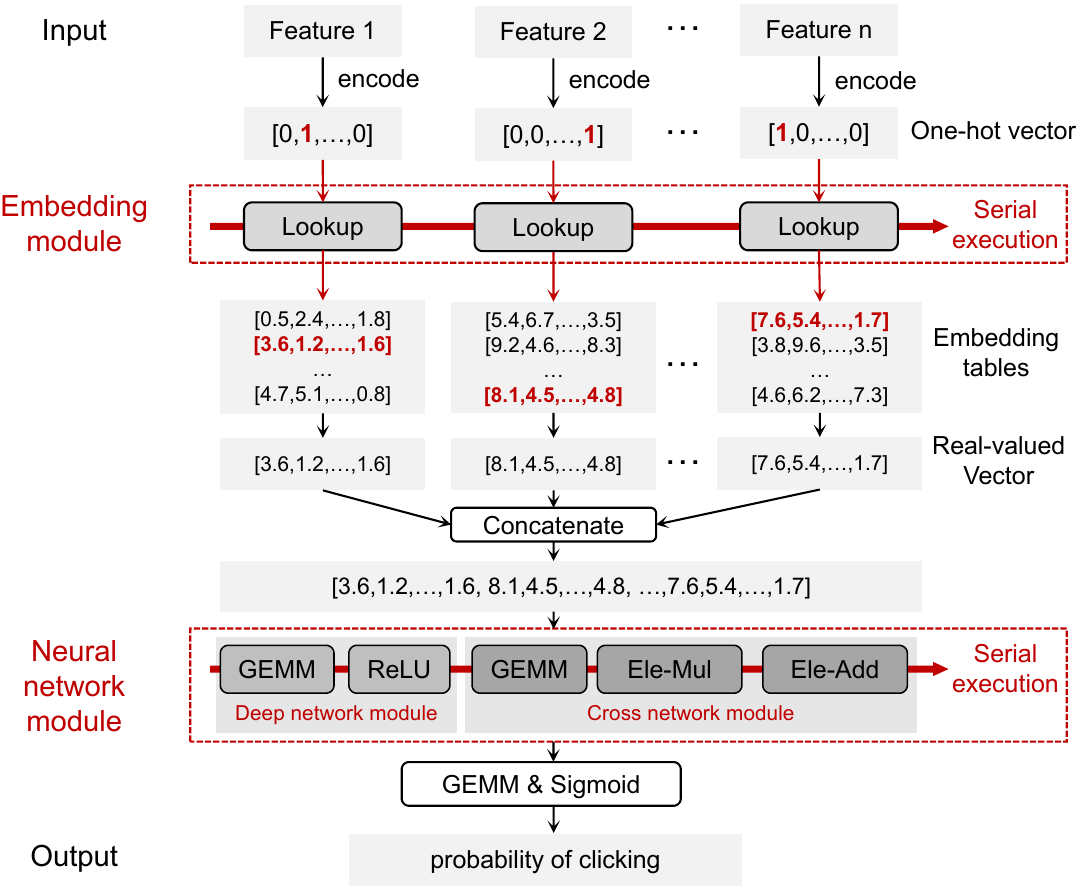}
  \caption{Take DCN~\cite{wang2017deep} as an example of CTR model inference. 
           General Matrix Multiplication is denoted as GEMM.}
  \label{fig: fig1}
\end{figure}

To meet the demands of large-scale and dynamic recommendation environments, Deep Learning Recommendation Models (DLRMs) have become the industry standard. 
As shown in Figure~\ref{fig: fig1}, a typical DLRM consists of two main components: the embedding module and the neural network module. 
The embedding module handles high-dimensional categorical features by projecting them into dense representations using large embedding tables. 
These embeddings are concatenated and fed into the neural network module for feature interaction and final CTR prediction.

%---------------------------
%% %---------------------------

The inference phase of recommender systems must prioritize low latency, as even a few milliseconds of delay can significantly impact user engagement and system throughput. 
Optimizing inference is therefore critical to enable ranking more candidate items within a fixed time window, leading to better recommendations and higher click-through rates. 
Beyond latency, the efficient utilization of computational resources is equally important. 
Embedding module often involves frequent and irregular memory accesses, which place substantial pressure on memory bandwidth. 
Simultaneously, the neural network module is computationally intensive and demands high parallel efficiency from modern hardware such as the GPU. 
Inefficient resource usage can lead to bottlenecks in production systems, increase infrastructure costs, and limit the scalability of the recommender service.
Overall, a high-quality recommender system must achieve strong performance in three key aspects: prediction accuracy, response latency, and efficient use of computational resources.

\subsection{CTR Model Inference}
For the sake of consistency, three terms are first clarified: model, module, and operator. 
The important modules, key processes, and structure are also illustrated.
\label{sec: 2.1}

\begin{itemize}[leftmargin=1.1em]
\item \makebox[4.5em][l]{\textbf{Model}} The complete model consisting of several modules, e.g., DCN~\cite{wang2017deep}, DCNv2~\cite{wang2021dcn}. 
\item \makebox[4.5em][l]{\textbf{Module}} The component including a certain number of operators in the model, e.g., Embedding, Cross Network. The module is the traditional smallest unit that can be executed in parallel by PyTorch on GPU.
\item \makebox[4.5em][l]{\textbf{Operator}} The function that performs computational tasks on the device, e.g., embedding lookup(), element-wise add(). The small computational operators can be fused into several single operators to be executed.
\end{itemize}

\noindent

With these concepts, the typical inference process of the CTR model can be described in three main steps, as shown in Figure~\ref{fig: fig1}.
First, the original sparse data is input into the embedding module for multiple table lookup operations, resulting in a dense vector of real values. 
Next, the concatenated dense vector is input to the neural network module, which performs more complex computations, such as matrix multiplication. 
Finally, the computed results are fed into a prediction layer to generate predictions. 
The roles of the two core modules are as follows: 

\textbf{Embedding Module} transforms sparse high-dimensional one-hot vectors of categorical features into dense low-dimensional real-valued vectors. 
Each feature field has a unique embedding table $\mathrm{E} \in \mathbb{R}^{n \times d}$, where $n$ is the number of features and $d$ is the embedding dimension. 
After performing a table lookup operation for each feature field separately, their vector representations are obtained. 
Then these embedding vectors of different feature fields are concatenated as input into neural network module.

\textbf{Neural Network Module} typically consists of two parallel modules: explicit interaction module (e.g., cross network in Figure~\ref{fig: fig1}) and implicit interaction module (e.g., deep network in Figure~\ref{fig: fig1}).
The two modules share the same input and have no data dependencies on each other and are computationally independent of each other. 
The cross network module learns explicit interaction features, and the deep neural network learns implicit interaction features. 
Finally, the outputs of these two modules are concatenated into a prediction layer to make predictions.

By analyzing the inference process, we can observe that the CTR model inherently possesses the parallel structure.
Firstly, all lookup operators in embedding module are independent of each other. 
They are each associated with their respective feature fields and do not query the same embedding table. 
Secondly, the computations in the explicit interaction module and the implicit interaction module are mutually independent.
They are executed without inter-module dependencies and can be performed in parallel without requiring intermediate results from each other.
This independence implies that these operations can be executed concurrently, providing the foundation for our parallel speedup.

\subsection{CUDA Stream Scheduling}
\label{sec: CUDA Stream}

%---------------------------
\begin{figure}
\centering
  \includegraphics[width=0.33\textwidth]{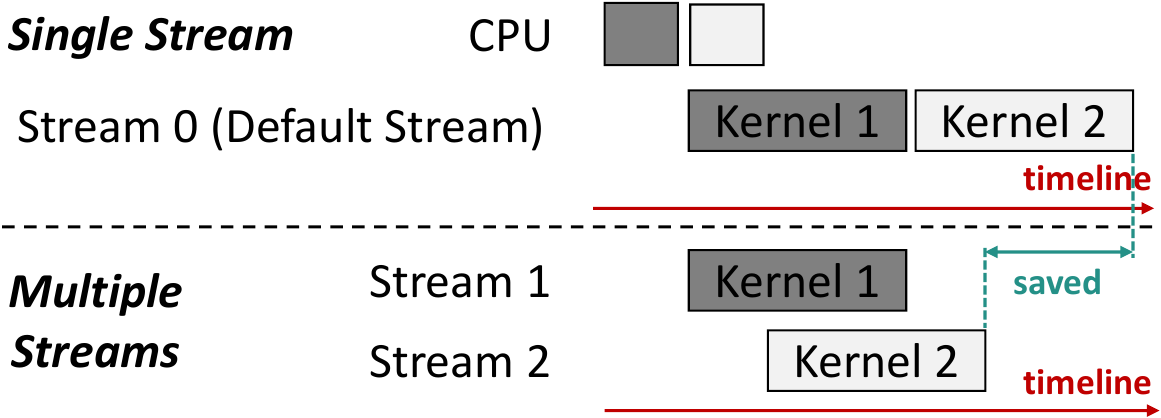}
  \caption{An example of CUDA streams.}
  \label{fig: fig2}
\end{figure}
%% %---------------------------

The stream is a core abstraction in the NVIDIA CUDA programming model, used to achieve asynchronous concurrent execution and task scheduling.
A stream refers to a sequence of operations delivered to a device for execution~\cite{luitjens2015cuda}. 
These operations include data transmission and computational tasks, also referred to as the aforementioned operators. 
The operators are fed into the stream.
When resources are available, the GPU executes them using the CUDA kernel from the stream. 
As shown in Figure~\ref{fig: fig2}, a single CUDA stream can contain multiple operations, which must be launched strictly in the order that they are added to the stream. 
Once sufficient resources are available, operations from multiple streams can be scheduled for concurrent execution~\cite{luitjens2015cuda}, with the default scheduling following the depth-first strategy.

Existing high-level frameworks, such as PyTorch, typically use only the default stream (``stream 0") for CTR model inference, meaning the operators are executed on the GPU in serial. 
Even if developers explicitly specify a stream in PyTorch, the execution may still not follow the developers' instructions precisely.
The limitation in CUDA stream scheduling, especially the reliance on a single default stream and the coarse-grained scheduling control in high-level frameworks, results in suboptimal parallelism and inefficiencies in CTR model inference on GPU.

\section{MOTIVATION}
\label{sec: Motivation}

Modern CTR prediction models have evolved toward increasingly complex architectures, yet their inference efficiency on GPU remains unsatisfactory. 
Through detailed analysis, we find that the root cause lies in the mismatch between the intrinsic characteristics of CTR models and the execution paradigm of GPU systems. 
In the following, we identify four key challenges and corresponding opportunities that motivate the design of DPIFrame.

\textbf{Challenge 1: Structural parallelism vs. serialized execution.}
CTR models inherently exhibit a high degree of parallelism at both the operator and module levels. 
For example, lookup operations in the embedding module are independent across feature fields, and the explicit and implicit interaction modules in the neural network component can be executed without interdependencies. 
However, existing deep learning frameworks typically rely on coarse-grained execution models and default CUDA stream, resulting in largely serialized operator execution. 
This prevents the model from leveraging its intrinsic parallel structure and leads to low GPU utilization. 
\textbf{Opportunity 1: This observation suggests that the inherent independence across operators and modules can be explicitly exploited to enable concurrent execution.}
By exposing and coordinating both intra-module and inter-module parallelism, it is possible to align the execution model with the parallel nature of CTR architectures and significantly improve hardware efficiency.

\textbf{Challenge 2: Irregular memory access vs. hardware efficiency.}
The embedding module introduces sparse and irregular memory access patterns due to large-scale table lookups. 
These accesses lack spatial and temporal locality, which conflicts with the design of GPU memory hierarchies and leads to inefficient bandwidth utilization. Consequently, embedding operations often become the dominant performance bottleneck in CTR inference. 
\textbf{Opportunity 2: Despite the irregularity of individual accesses, the overall embedding workload is deterministic and known in advance during inference.}
This enables a global view of memory access patterns and provides an opportunity to reorganize embedding operations into a structured computation process. 
By transforming scattered lookups into coordinated, sequentialized memory accesses, we can significantly improve memory efficiency and alleviate bandwidth bottlenecks.

\textbf{Challenge 3: Fine-grained operators vs. GPU execution efficiency.}
CTR model inference is dominated by numerous lightweight operators such as element-wise vector computations. 
These operators typically have very short execution times, while the overhead of launching CUDA kernels is comparatively high. 
As a result, a significant portion of execution time is spent on kernel invocation rather than actual computation, leading to poor overall efficiency. 
Moreover, the low computational load of these operators makes it difficult to directly parallelize them to effectively overlap execution latency. 
\textbf{Opportunity 3: To address this inefficiency, it is critical to increase computational granularity.}
By fusing multiple small operators and reorganizing workloads into larger computational units, we can reduce kernel launch overhead, improve data locality, and better match GPU execution characteristics. 
Such transformations enable more efficient utilization of GPU compute resources and create sufficiently large workloads that are amenable to parallel execution and latency hiding.

%---------------------------
\begin{figure}
\centering
  \includegraphics[width=0.36\textwidth]{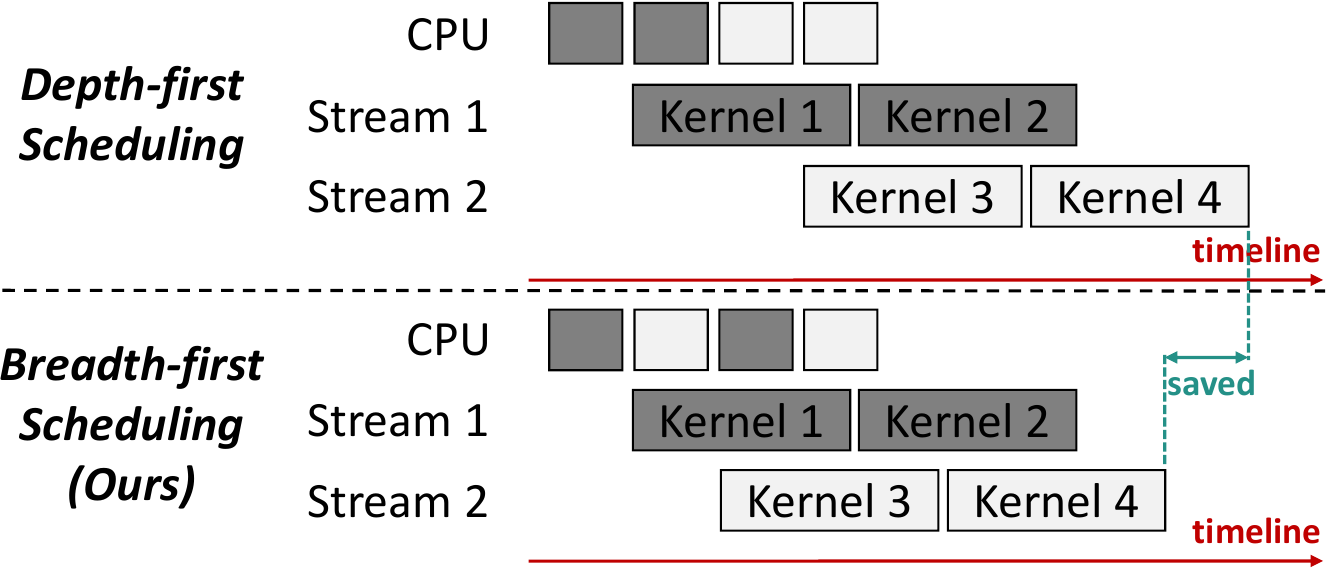}
  \caption{Depth-first and breadth-first scheduling. 
  Depth-first scheduling prioritizes launching all operators within a CUDA stream before initiating operators in another CUDA stream. 
  Breadth-first scheduling entails interleaving the launch of operators from different CUDA streams.}
  \label{fig: fig3}
\end{figure}
%% %---------------------------

\textbf{Challenge 4: Coarse-grained scheduling vs. fine-grained parallelism.}
Existing frameworks typically employ coarse-grained scheduling strategies at the module level, often following depth-first execution orders. 
Such strategies fail to fully exploit the fine-grained parallelism present in CTR models and delay the execution of independent operators that could otherwise run concurrently. 
As a result, many GPU resources remain underutilized due to insufficient overlap between operations. 
\textbf{Opportunity 4: This limitation highlights the need for a more fine-grained scheduling mechanism.}
By introducing operator-level scheduling across multiple CUDA streams and carefully coordinating their execution order, it is possible to maximize concurrency and overlap computation effectively. 
In particular, as shown in Figure~\ref{fig: fig3}, adopting a breadth-first scheduling strategy allows independent operators from different modules to be launched earlier, thereby improving overall execution efficiency and GPU utilization.

In summary, the inefficiency of CTR model inference arises from the mismatch between fine-grained, parallel model structures and coarse-grained execution mechanisms. 
These challenges motivate a systematic redesign that jointly considers parallelism extraction, workload restructuring, and execution scheduling. 
Based on these insights, we propose DPIFrame, which leverages dual-level parallelism and optimized scheduling to bridge this gap and achieve efficient CTR model inference on GPU.

\section{DPIFRAME DESIGN}
\label{sec: DPISys Design}
%-------------------------------------------------------------------------------

\begin{figure}
\centering
  \includegraphics[width=1.0\linewidth]{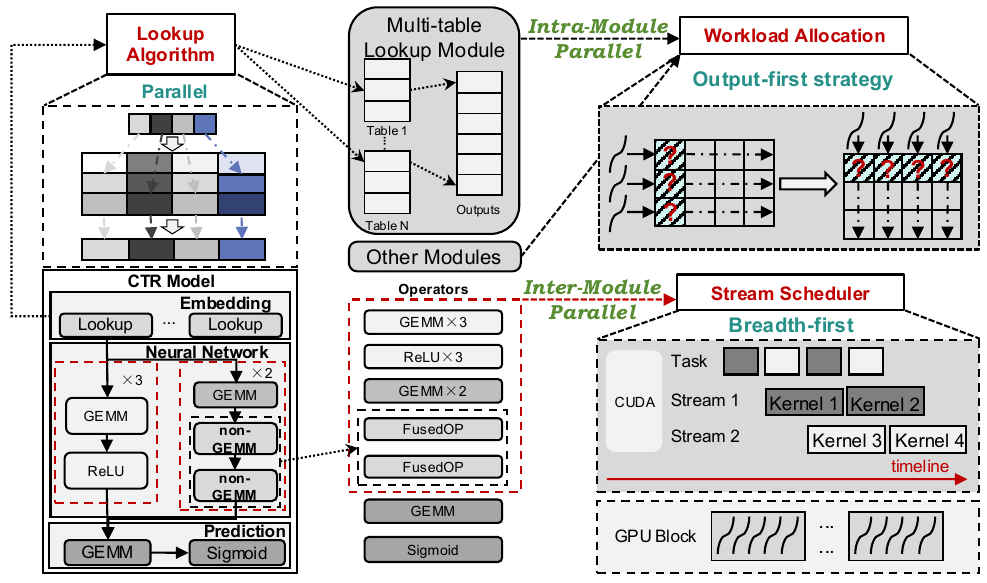}
  \caption{The overview of \ours.}
  \label{fig: overview}
\end{figure}

As shown in Figure~\ref{fig: overview}, \ours, as the first dual parallel execution inference system for CTR prediction, supports both \textbf{intra-module} and \textbf{inter-module} parallel execution. 
In this section, we first present the overall architecture of \ours.
We then outline the core parallel design following the CTR model inference process.
In the embedding stage, we design the multi-table lookup algorithm and focus on workload allocation.
In the neural network stage, we describe the fusion of non-GEMM operators and the corresponding stream scheduling strategy.

\subsection{Dual Parallel Architecture}
%-------------------------------------------------------------------------------

As aforementioned, we identify the critical limitation of the CTR model inference inefficiency as the mismatch between the serial computational pattern and the parallel model structure. 
Furthermore, we capture that the computational load of the operators in the CTR model is short tasks, making it difficult to directly parallelize them to overlap the latency. 
Therefore, we propose a dual parallelizable architecture of the CTR model inference system that performs parallel CTR model inference both intra-module and inter-module. 
We also process the non-GEMM operators in a module by merging them into a few ones, then their workloads are intensively allocated to threads, thus enabling our multi-table parallel lookup algorithm and achieving intra-module parallelism. 
Besides, the operators of parallel modules are alternately scheduled to different CUDA streams to maximize overlap, thereby boosting the inter-module parallelism.

\subsection{Multi-table Lookup Algorithm}
\label{sec: Multi-table}
%-----------------------------------
In the embedding stage, we design the multi-table lookup algorithm that leverages the independence of lookup operations to unleash the parallel potential in the model structure.
Suppose the input $\mathrm{X}$ of the CTR model is of size $b \times k$, where $b$ represents the number of samples and $k$ represents the number of feature fields. 
Each feature field has a separate embedding vector table $\mathrm{E}_{i} \in \mathbb{R}^{n_{i} \times d}$, where $i$ represents $i$-th feature field, $n_{i}$ represents the number of features contained in the $i$-th feature field. 
For example, the "gender" feature field contains male and female features, so the $n$ of this feature field is 2. 
Some feature fields have a huge number of features, which can reach millions or even tens of millions, so their corresponding embedding vector table is also very large.
$d$ represents the dimension of the embedding table, that's the length of the vector used to represent each feature. 
For the input samples, the $k$ feature fields are first encoded with one-hot encoding, respectively. 
To reduce sparsity and memory consumption, they are often further represented as numeric IDs. 

\begin{figure}
\centering
  \includegraphics[width=0.7\linewidth]{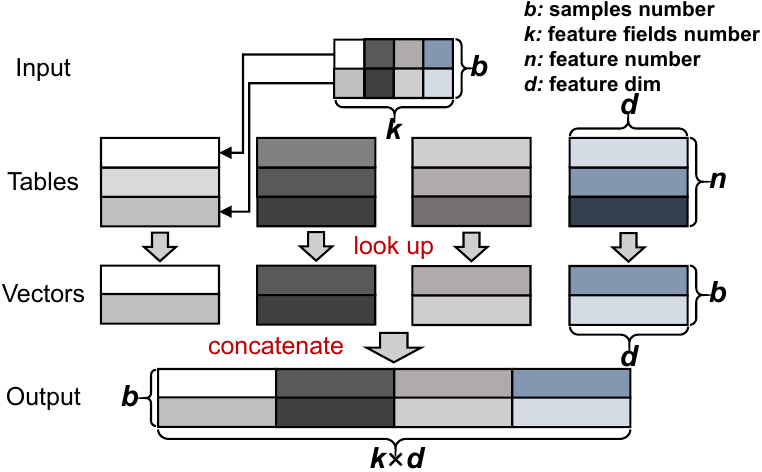}
  \caption{An example of embedding table lookups. Where the number of samples is 2, the number of feature fields is 4, and the number of features per feature field is 3.}
  \label{fig: fig5}
\end{figure}

An example of embedding operations is shown in Figure~\ref{fig: fig5}, which looks up its corresponding embedding table for each of these $k$ feature fields. 
After a total of $k$ lookup operations to obtain their vector representations. 
Then, these vectors are concatenated into a matrix $\mathrm{X}_{embed} \in \mathbb{R}^{b \times (k \times d)}$ as the input to the next layer of the model.
All the embedding lookups are as a whole, so it is entirely possible to replace the original $k$ serial lookups with a single parallel operation.
We can compute the size of the embedded output based on the embedding configuration parameters as well as the size of the input. 
Therefore, we only need to initialize the output matrix and populate it element by element, without any additional intermediate storage vectors. 

\begin{algorithm}[t]
\footnotesize
\caption{Multi-table Lookup}
\label{alg: alg1}
\DontPrintSemicolon
\SetCommentSty{mycommfont}

\KwIn{
\begin{tabular}[t]{@{}l@{}}
$IDs \in \mathbb{N}^{b \times k}$: Input feature IDs \\[1pt]
$Embs$: Embedding tables, where the $i$-th table is of size $n_i \times d$
\end{tabular}
}

\KwOut{
$EmbedOut \in \mathbb{R}^{b \times (k \times d)}$: Lookup results
}

\BlankLine

$total\_elements \gets b \times k \times d$\;
$row\_width \gets k \times d$\;

\For{$idx \gets 0$ \KwTo $total\_elements - 1$}{

    $row \gets \left\lfloor \dfrac{idx}{row\_width} \right\rfloor$\;
    
    $col \gets idx \bmod row\_width$\;

    $table\_id \gets \left\lfloor \dfrac{col}{d} \right\rfloor$\;

    $emb\_row \gets IDs[row \times k + table\_id]$\;

    $emb\_col \gets col \bmod d$\;

    $EmbedOut[idx] \gets Embs[table\_id][emb\_row \times d + emb\_col]$\;
}

\KwRet{$EmbedOut$}

\end{algorithm}

To match each position of the matrix with the corresponding position of its embedding table, we design Algorithm~\ref{alg: alg1} to accomplish the global embedding table lookup. 
Firstly, we compute the size of the output matrix as well as the number of its columns based on the size of the input and the embedding configuration parameters (line 1-2). 
Then start traversing the output matrix. 
Next, iterate through the matrix row by row (line~3), based on the step, derive the position of the current element in terms of the row and column to which it belongs in the matrix (line 4-5). 
Further, we can compute to know which embedding table to look up (line 6). 
Finally, leveraging the gathered information along with the input IDs, we determine the specific row and column in the embedding table (lines 7-8), allowing us to find the value to be filled in the currently traversed output matrix element (line 9).

The optimization design consolidates multiple table lookups across different feature fields into a single operation. 
The algorithm only requires very few computations to derive the indices. 
To face multi-hot encoded features, it can also support sequential features by marking offset information.

\subsection{Workload Allocation}
\label{sec: Workload Allocation}
%-----------------------------------

\begin{figure}
\centering
  \includegraphics[width=0.8\linewidth]{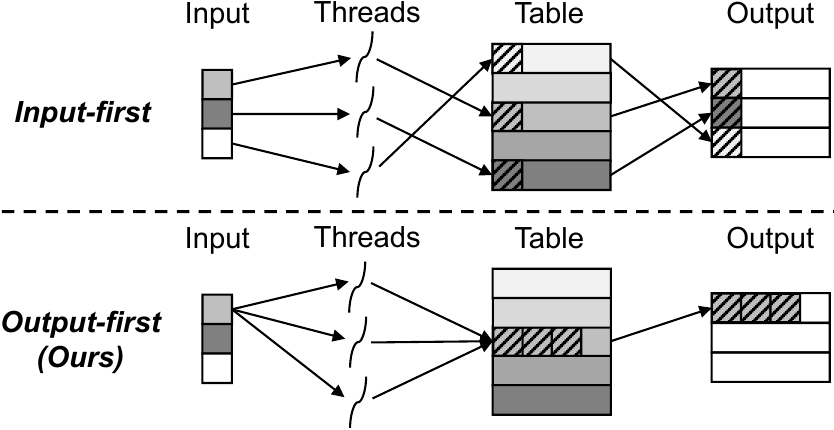}
  \caption{Output-first workload allocation's advantages on fused embedding lookup operation.}
  \label{fig: fig6}
\end{figure}

Once the total workload of fused operators is determined, the focus shifts to allocating the workload across threads, aiming at ensuring that it is GPU-friendly and further enhancing intra-module parallel execution. 
We design the output-first workload allocation methods, and take the fused embedding lookup operator as an example to explain its advantages, as shown in Figure~\ref{fig: fig6}.
As a comparison, we also introduce the drawbacks of the input-first approach.
The input-first scheme involves maximizing the allocation of different input samples among threads, while our output-first scheme involves coordinating threads to process the same sample together. 

According to the NVIDIA GPU programming model~\cite{guide2020cuda}, the smallest hardware execution unit on a GPU is a warp, which typically consists of 32 threads. 
When the 32 threads execute the same instruction, they operate in a Single Instruction, Multiple Threads (SIMT) fashion, allowing simultaneous reads and writes. 
When adopting an input-first scheme where different threads prioritize different samples, even though the input reads have a sequentially addressed pattern, the table lookups and output writes will involve random reads and writes instead of sequentially addressed ones.

This is hardware-unfriendly because the memory bandwidth is not fully utilized, resulting in wasted resources and low performance. 
Our output-first approach enables different threads to share the same sample and can perform embedding table lookups feature by feature. 
As a result, both input reads and table lookups, as well as output writes, remain sequentially addressed. 
This ensures optimal utilization of the hardware memory bandwidth and brings performance improvements.

\subsection{Fusion Mechanism for non-GEMM Operators}
\label{sec: Fusion Mechanism}

Different from embedding, neural network includes more varied operators. 
There are two typical modules in neural network: an implicit feature interaction module and an explicit feature interaction module. 
The implicit interaction module remains relatively consistent and typically consists of several layers of multi-layer perceptrons (MLPs), which contain GEMM and activation operators. 
But due to different design objectives, the explicit interaction modules of different deep CTR models often differ. 
For example, the explicit interaction module of DCNv2~\cite{wang2021dcn} is a cross network, which contains element-wise addition and element-wise multiplication operators besides GEMM. 
The explicit interaction module of DeepFM~\cite{guo2017deepfm} consists of a logistic regression and a factorization machine, which contain various operators such as table lookup, element-wise square and reduce-sum.

\begin{algorithm}[t]
\footnotesize
\caption{Breadth-first Stream Scheduling}
\label{alg: alg3}
\DontPrintSemicolon
\SetCommentSty{mycommfont}
\SetAlgoLined

\KwIn{
\begin{tabular}[t]{@{}l@{}}
$Explicit$: Explicit Interaction Module \\
$Implicit$: Implicit Interaction Module
\end{tabular}
}

\KwOut{
\begin{tabular}[t]{@{}l@{}}
$S_{explicit}$: CUDA stream for $Explicit$ \\
$S_{implicit}$: CUDA stream for $Implicit$ \\
$Q$: Breadth-first scheduling queue
\end{tabular}
}

\BlankLine

$Ops_{explicit} \gets Explicit.\mathrm{operators}()$\;
$Ops_{implicit} \gets Implicit.\mathrm{operators}()$\;

$n_{explicit} \gets |Ops_{explicit}|$\;
$n_{implicit} \gets |Ops_{implicit}|$\;

$S_{explicit} \gets \mathrm{cudaStreamCreate}()$\;
$S_{implicit} \gets \mathrm{cudaStreamCreate}()$\;

$S_{explicit}.\mathrm{Add}(Ops_{explicit})$\;
$S_{implicit}.\mathrm{Add}(Ops_{implicit})$\;

\BlankLine

\If{$n_{implicit} > n_{explicit}$}{
  \For{$i \gets 0$ \KwTo $n_{explicit} - 1$}{
       $Q.\mathrm{Push}(Ops_{implicit}[i])$\;
       $Q.\mathrm{Push}(Ops_{explicit}[i])$\;
   }
   \For{$j \gets n_{explicit}$ \KwTo $n_{implicit} - 1$}{
       $Q.\mathrm{Push}(Ops_{implicit}[j])$\;
   }
}
\Else{
   \For{$i \gets 0$ \KwTo $n_{implicit} - 1$}{
       $Q.\mathrm{Push}(Ops_{explicit}[i])$\;
       $Q.\mathrm{Push}(Ops_{implicit}[i])$\;
   }
   \For{$j \gets n_{implicit}$ \KwTo $n_{explicit} - 1$}{
       $Q.\mathrm{Push}(Ops_{explicit}[j])$\;
   }
}

\BlankLine

\KwRet{$S_{explicit},\ S_{implicit},\ Q$}

\end{algorithm}

To address the variety of neural network among CTR models, we propose a uniform fusion mechanism: we combine all neighbouring non-GEMM operators into a single operator. 
DPIFrame represents the model forward propagation by constructing a directed acyclic graph, in which nodes are operators and edges are tensors. 
Starting from the root node, we traverse the graph to mark all non-GEMM nodes connected by edges. 
We define the set of these nodes and the edges connecting them as a subgraph. 
Within a subgraph, we fuse the operators into a new operator. 
Moreover, the workload allocation follows a similar principle as described in Section \ref{sec: Workload Allocation}. 
We traverse the output matrix element by element, assigning the filling tasks for each matrix element to individual threads, ensuring that data reads and writes are contiguous in memory.

\subsection{Stream Scheduling Strategy}
\label{sec: Stream Scheduling Strategy}
%-----------------------------------

Beyond intra-module parallelism, when deploying models on the GPU, the opportunity for parallel execution exists for operators that are placed into different CUDA streams. 
However, existing deep learning frameworks typically use only one default stream to schedule operators. 
Operators are executed sequentially according to the scheduling order, even for models with parallel structures.

To achieve inter-module parallel execution and further address the mismatch between hardware execution and model structure, we design an operator-level stream scheduling strategy to fine-tune the parallel computation of the key modules in CTR models. 
The core idea of the stream scheduling strategy is to assign the operators of key modules to separate CUDA streams using a breadth-first scheduling approach.
The operators from different modules are added to different CUDA streams. 
The operators are then pushed into the launch queue in an alternating pattern by module wise. 
This can ensure operators are executed on GPU as early as possible to avoid blocking or waiting.
The module that has more operators launches first in our \ours, it can help hide the startup costs of many small computation tasks.
The specific design details are described in Algorithm~\ref{alg: alg3}.

\section{EVALUATION}
\label{sec: Evaluation}
%-------------------------------------------------------------------------------

In this section, we evaluate \ours~through a set of comprehensive experiments. 
We introduce the experimental setup (\ref{sec: Experimental Setup}), to verify that \ours~meets the requirements for accuracy (\ref{sec: Prediction Accuracy}) and efficiency (\ref{sec: Speedup}), analyze each key design component of \ours~improvement(\ref{sec: Breakdown Analysis}), evaluate the hardware utilization improvements(\ref{sec: Hardware Utilization}),
We also analyze the embedding sensitivity, the advantages of the designed steam scheduling strategy, and point out an interesting phenomenon observed during multi-stream scheduling, aiming to inspire further research.
Since embedding is an essential module of all CTR models, we focus on embedding sensitivity studies to ensure that \ours~delivers consistent performance gains under various complex configurations (\ref{sec: Embedding Sensitivity Studies}). 
We then evaluate the advantages of the designed stream scheduling strategy compared to single-stream and general multi-stream scheduling (\ref{sec: Evaluation for Scheduling Strategy}). 
Finally, we test different startup sequences for the configuration, aiming to achieve a comprehensive assessment (\ref{sec: Discussion}). 

\subsection{Experimental Setup}
\label{sec: Experimental Setup}
%-----------------------------------

\textbf{Hardware.} We deploy \ours~on a server equipped with an Intel Core i9-10980XE CPU (3.0 GHz, 18 Cores) and 256GB DRAM in total, and a single NVIDIA RTX 3090 GPU with 24GB DRAM.

\textbf{Software.} The original implementation of CTR models is based on an open-source project, FuxiCTR~\footnote{\url{https://github.com/reczoo/FuxiCTR}}. 
We implemented \ours~as a building block and embedded it into PyTorch (v1.13.0) using pybind11. 
The environment we used includes CUDA 11.6, and the version of gcc is 9.4.0.

\textbf{Datasets.} To thoroughly evaluate the accuracy and efficiency of \ours, while ensuring its effectiveness in real-world scenarios, we conducted experiments on two real-world benchmark datasets released by internationally renowned advertising companies: Avazu~\footnote{\url{https://www.kaggle.com/c/avazu-ctr-prediction}} and Criteo~\footnote{\url{https://www.kaggle.com/c/criteo-display-ad-challenge}}. 
They contain 24 and 39 feature fields, respectively, with significant variations in the number of features within each field, reflecting real-world scenarios. 

\textbf{Models.} We validate \ours~using four industry-recognized CTR models: DCN~\cite{wang2017deep}, DCNv2~\cite{wang2021dcn}, Wide\&Deep~\cite{cheng2016wide}, and DeepFM~\cite{guo2017deepfm}. 
These models have been validated in real application scenarios, and their diverse structures enable a comprehensive evaluation of DPIFrame designs.

\textbf{Configuration parameter.} To validate that \ours~provides effective acceleration in different scenarios, our experiments used various model configuration parameters. 
Embedding dimension: 16 and 32. 
Hidden size units of deep network module: [256, 256, 256], [512, 512, 512], and [1024, 1024, 1024]. 
Batch size is typically set to 2048. 
For DCN~\cite{wang2017deep} and DCNv2~\cite{wang2021dcn}, the number of cross layers is 3. 
In summary, each model is validated under 6 sets of configuration parameters.

\textbf{Metrics.} We evaluate \ours~from the following metrics: 
(1) AUC and LogLoss. 
These are two classic metrics for evaluating the accuracy of CTR models~\cite{guo2017deepfm}. 
The premise of \ours~acceleration is to rigorously ensure the accuracy of the model. 
(2) Speedup and latency. 
The main goal of \ours~is to accelerate the inference of CTR models. 
(3) Hardware utilization. 
\ours~pursues higher hardware utilization to achieve fast model inference. 

\textbf{Baselines.} We compare \ours~with four strong frameworks: PyTorch~\cite{paszke2019pytorch}, TorchRec~\cite{ivchenko2022torchrec}, HugeCTR~\cite{wang2022merlin}, and OneFlow~\cite{yuan2021oneflow}. 
TorchRec is a PyTorch-based domain library for recommendation systems. 
It enables researchers to build state-of-the-art personalized models and deploy them in production environments. 
TorchRec includes optimized recommender system kernels driven by FBGEMM~\cite{khudia2021fbgemm}, which enables high-speed GPU inference. 
Currently, TorchRec is used in Meta's recommendation system platform. 
HugeCTR is an open-source GPU-accelerated integration framework developed by NVIDIA. 
It is specifically designed to optimize both the training and inference processes for CTR models. 
The OneFlow team has introduced an efficient, scalable, and highly flexible recommendation system component called OneEmbedding~\footnote{\url{https://docs.oneflow.org/en/master/cookies/one\_embedding.html}}. 
Its performance exceeds that of typical frameworks and even outperforms the dedicated recommendation framework developed by NVIDIA, HugeCTR.

\begin{figure*}
  \centering
  \subfigure[Speedup on the Avazu Dataset]{\label{fig: speedup_avazu}\includegraphics[width=1.0\textwidth]{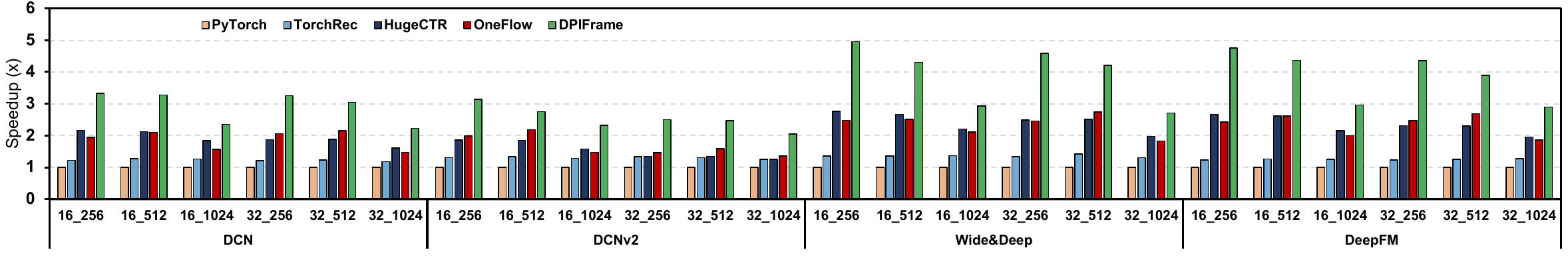}}
  \subfigure[Speedup on the Criteo Dataset]{\label{fig: speedup_ctireo}\includegraphics[width=1.0\textwidth]{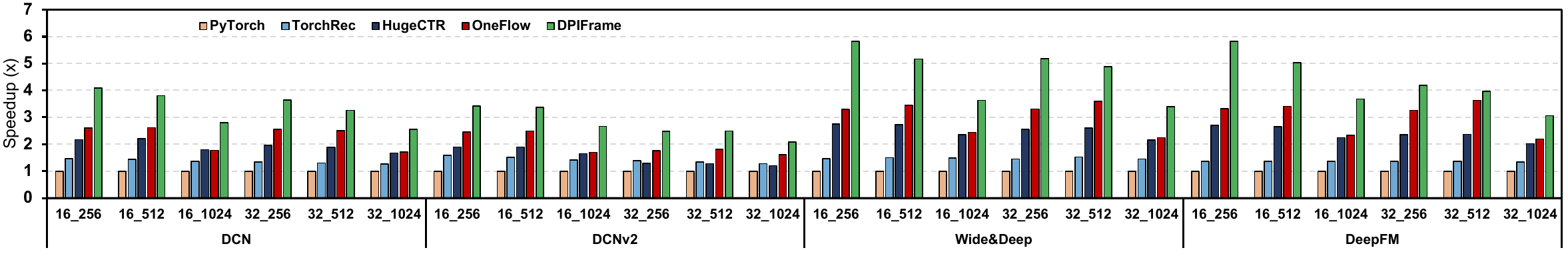}}
  \caption{The overall speedup comparison and \ours~on four models with different computational scales.}
  \label{fig: overall speedup}
\end{figure*}

\begin{table}[t]
\centering
\renewcommand\arraystretch{1.1}
\caption{Prediction accuracy of \ours.}
\begin{tabularx}{\columnwidth}{c|
  >{\centering\arraybackslash}X
  >{\centering\arraybackslash}X|
  >{\centering\arraybackslash}X
  >{\centering\arraybackslash}X}
\hline
\multirow{2}{*}{\textbf{-}} & \multicolumn{2}{c|}{\textbf{Avazu}} & \multicolumn{2}{c}{\textbf{Criteo}} \\ \cline{2-5} 
                  & \textbf{AUC}         & \textbf{LogLoss}      & \textbf{AUC}         & \textbf{LogLoss}      \\ \hline
\textbf{DCN}               & 0.7922      & 0.3725       & 0.8094      & 0.4422       \\ \hline
\textbf{DCNv2}             & 0.7935      & 0.3718       & 0.8113      & 0.4405       \\ \hline
\textbf{Wide}\textbf{\&}\textbf{Deep}        & 0.7925      & 0.3723       & 0.8089      & 0.4432       \\ \hline
\textbf{DeepFM}            & 0.7928      & 0.3722       & 0.8061      & 0.4461       \\ \hline
\end{tabularx}
\label{table: accuracy}
\end{table}

%-------------------------------------------------------------------------------

\subsection{Prediction Accuracy}
\label{sec: Prediction Accuracy}
%-----------------------------------

We train CTR models on two datasets using PyTorch and evaluate the accuracy of these models on test sets when performing inference using both PyTorch and \ours. 
The purpose of this evaluation is to verify that \ours strictly maintains model accuracy. 
The specific performance is shown in Table~\ref{table: accuracy}.
The experimental results show that for both the AUC and LogLoss metrics, \textbf{\ours~and PyTorch achieve identical values to the fourth decimal place (in fact, they are identical to the sixth decimal place)}. 
This shows that the speedup achieved by \ours~can be trusted. 

%-----------------------------------

\subsection{End-to-End Speedup}
\label{sec: Speedup}
%-----------------------------------

We evaluate the end-to-end speedup of \ours~with PyTorch, TorchRec, HugeCTR, and OneFlow on four CTR models with six sets of configuration parameters on two datasets, as shown in Figure~\ref{fig: overall speedup}. 
X\_Y means embedding dimension and hidden layer size of MLP separately. 
The experimental results show that \ours~comprehensively outperforms the comparative baselines. 

\textbf{Framework-wise comparison.} 
The results show that \ours~achieves average speedups of 3.55$\times$, 2.63$\times$, 1.69$\times$, and 1.53$\times$ compared to PyTorch, TorchRec, HugeCTR, and OneFlow, with maximum speedups of \textbf{5.83$\times$, 4.29$\times$, 2.15$\times$, and 2.0$\times$}, respectively. 
\ours~can significantly improve the performance of inference with the help of intra-module and inter-module parallel execution.

\begin{table}[t]
\caption{Speedup of \ours~compared to baselines on four models.}
\centering
\renewcommand\arraystretch{1.1}
\scalebox{0.8}{\begin{tabular}{c|cccccccc}
\hline
\multirow{3}{*}{\textbf{-}} & \multicolumn{8}{c}{\textbf{\ours~VS.}}                                                                                                      \\ \cline{2-9} 
                  & \multicolumn{2}{c|}{\textbf{PyTorch}}     & \multicolumn{2}{c|}{\textbf{TorchRec}}    & \multicolumn{2}{c|}{\textbf{HugeCTR}}     & \multicolumn{2}{c}{\textbf{OneFlow}} \\ \cline{2-9} 
                  & Avg. & \multicolumn{1}{c|}{Max.} & Avg. & \multicolumn{1}{c|}{Max.} & Avg. & \multicolumn{1}{c|}{Max.} & Avg.         & Max.         \\ \hline
\textbf{DCN}               & 3.14$\times$ & \multicolumn{1}{c|}{4.08$\times$} & 2.41$\times$ & \multicolumn{1}{c|}{2.77$\times$} & 1.62$\times$ & \multicolumn{1}{c|}{1.88$\times$} & 1.51$\times$         & 1.71$\times$         \\ \hline
\textbf{DCNv2}             & 2.65$\times$ & \multicolumn{1}{c|}{3.43$\times$} & 1.94$\times$ & \multicolumn{1}{c|}{2.41$\times$} & 1.73$\times$ & \multicolumn{1}{c|}{1.95$\times$} & 1.46$\times$         & 1.71$\times$         \\ \hline
\textbf{Wide\&Deep}        & 4.32$\times$ & \multicolumn{1}{c|}{5.83$\times$} & 3.03$\times$ & \multicolumn{1}{c|}{3.94$\times$} & 1.72$\times$ & \multicolumn{1}{c|}{2.12$\times$} & 1.60$\times$         & 2.00$\times$         \\ \hline
\textbf{DeepFM}            & 4.08$\times$ & \multicolumn{1}{c|}{5.82$\times$} & 3.13$\times$ & \multicolumn{1}{c|}{4.29$\times$} & 1.71$\times$ & \multicolumn{1}{c|}{2.15$\times$} & 1.54$\times$         & 1.96$\times$         \\ \hline
\end{tabular}}
\label{table: model-wise speedup}
\end{table}

\textbf{Dataset-wise comparison.} 
As shown in Figure~\ref{fig: speedup_avazu}, on the Avazu dataset, \ours~achieves average speedups of 3.32$\times$, 2.59$\times$, 1.61$\times$, and 1.60$\times$ compared to PyTorch, TorchRec, HugeCTR, and OneFlow, with maximum speedups of 4.95$\times$, 3.89$\times$, 1.88$\times$, and 2.0$\times$, respectively. 
Figure~\ref{fig: speedup_ctireo} shows on the Criteo dataset, \ours~achieves average speedups of 3.77$\times$, 2.67$\times$, 1.78$\times$, and 1.46$\times$ compared to PyTorch, TorchRec, HugeCTR, and OneFlow, with maximum speedups of 5.83$\times$, 4.29$\times$, 2.15$\times$, and 1.77$\times$, respectively. 
The acceleration performance achieved on these two datasets is similar, with slightly better performance on the Criteo dataset compared to the Avazu dataset. 
This observation underscores \ours's ability to deliver excellent performance even in large data scenarios.

\textbf{Model-wise comparison.} 
The results of the speedup for the different models are shown in Table~\ref{table: model-wise speedup}. 
Overall, the acceleration achieved for Wide\&Deep and DeepFM is slightly higher than for DCN and DCNv2. 
The reason is that, in Wide\&Deep and DeepFM, their feature interaction modules primarily involve embedding table lookup operations. 
This observation further validates the effectiveness of \ours's intra-module parallelism optimization for embedding. 
Compared to HugeCTR and OneFlow, the performance improvement achieved for the DeepFM model is relatively modest. 
This is because both HugeCTR and OneFlow implement specific optimizations for the inner product units of DeepFM.

\begin{figure*}
\centering
  \includegraphics[width=1.0\textwidth]{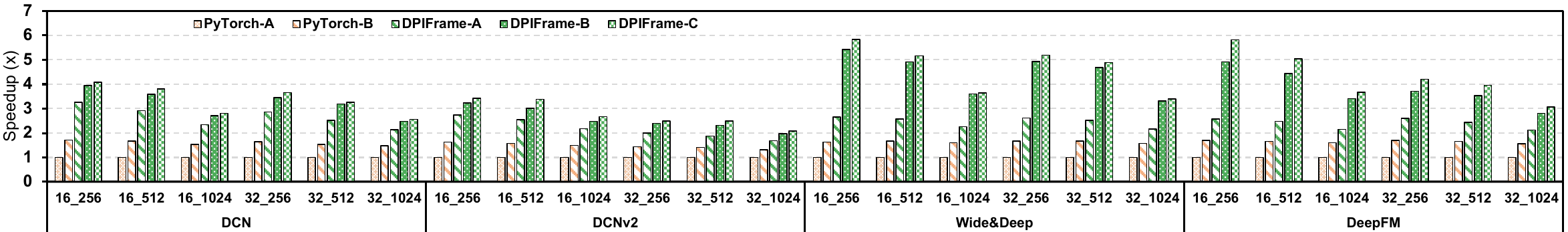}
  \caption{The performance breakdown by components of \ours~on four models with different computational scales.}
  \label{fig: breakdown}
\end{figure*}

\subsection{Breakdown Analysis}
\label{sec: Breakdown Analysis}

We next break down \ours~performance by designing components. 
The evaluation is conducted based on the Criteo dataset as shown in Figure~\ref{fig: breakdown}. 
We chose PyTorch as a baseline and compared the performance gains brought by each core component of \ours. 
PyTorch-A represents the original, non-optimized implementation of FuxiCTR. 
We observed that in the data preprocessing phase, it transfers different feature fields separately to the GPU. 
When each feature field undergoes an embedding table lookup, it requires a data type conversion (from float to long) each time, which wastes time. 
Therefore, we consolidate the feature field transfers into a single operation, and the data type conversions are also consolidated into a single operation, resulting in PyTorch-B. 
This allows us to more clearly demonstrate the optimization performance brought by the components of \ours. 
\ours-A represents only performing intra-module parallel execution for embedding. 
\ours-B introduces intra-module parallel execution for neural network on top of \ours-A. 
\ours-C implements inter-module parallel execution on top of \ours-B. 

The results show that compared to PyTorch-A and PyTorch-B, \ours-A achieves average speedups of 2.42$\times$ and 1.52$\times$, with maximum speedups of 3.27$\times$ and 1.91$\times$, respectively. 
This clearly validates the effectiveness of \ours's embedding multi-table parallel lookup algorithm. 
Building on \ours-A, \ours-B achieves an average performance improvement of 28.0\% and a maximum improvement of 50.9\%. 
This highlights the benefits of \ours's non-GEMM operator fusion mechanism. 
Finally, based on \ours-B, \ours-C further exploits the GPU's capabilities, achieving an average gain of 6.3\% and a maximum gain of 15.4\%. 
The performance gains from scheduling multiple CUDA streams are modest because when multiple tasks occupy a finite number of compute units, the execution time of each task becomes longer.

%-------------------------------------------------------------------------------

\begin{figure}
  \centering
  \subfigure[Avazu]{\includegraphics[width=0.48\linewidth]{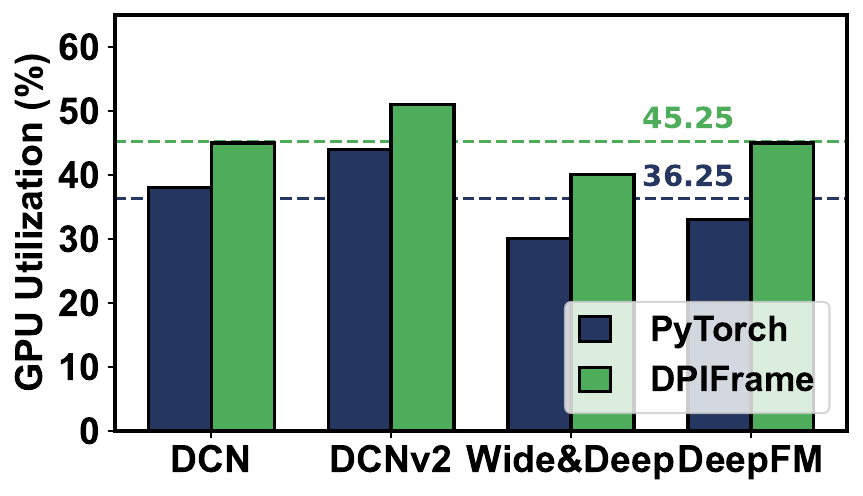}}
  % \hfill
  \subfigure[Criteo]{\includegraphics[width=0.48\linewidth]{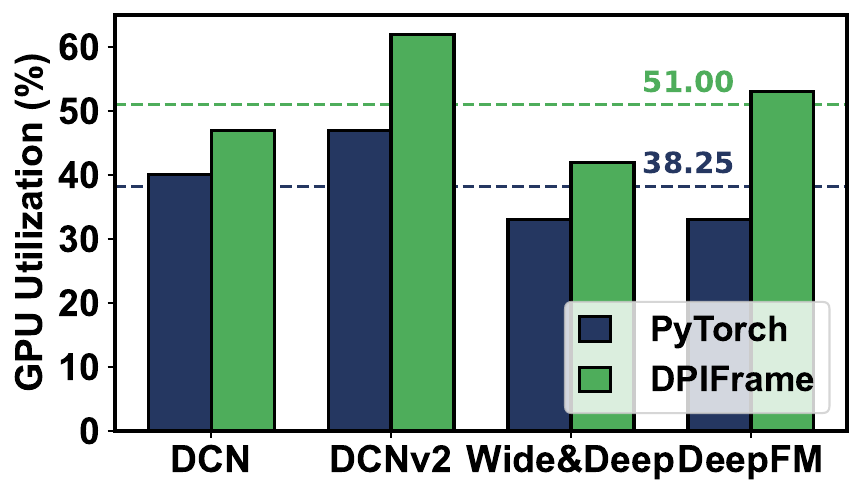}}
  \caption{GPU utilization comparison between PyTorch and DPIFrame on the Avazu and Criteo datasets. DPIFrame consistently achieves higher GPU utilization across all evaluated CTR models, demonstrating its ability to better exploit GPU resources through optimized operator scheduling and execution.}
  \label{fig: GPU Utilization}
\end{figure}

\begin{figure*}
\centering
  \includegraphics[width=1.0\textwidth]{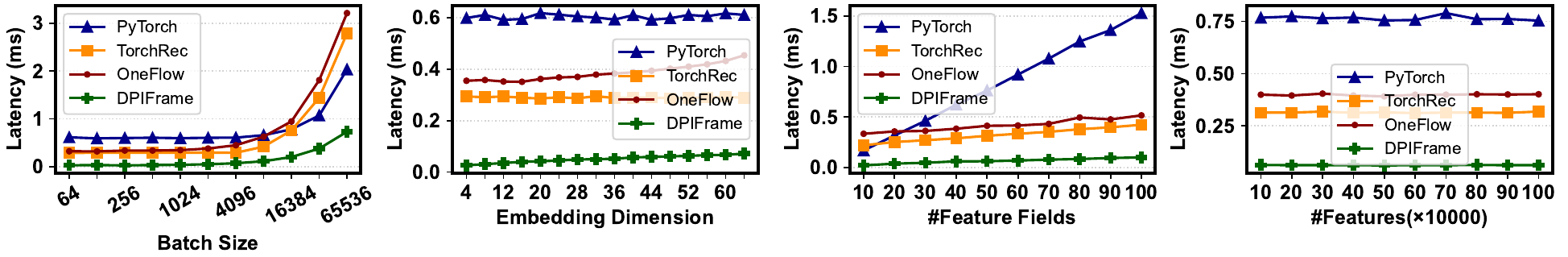}
  \caption{The embedding performance comparison between baselines and \ours~in different scenarios.}
  \label{fig: emb_sensitivity}
\end{figure*}

\subsection{Hardware Utilization}
\label{sec: Hardware Utilization}
%-----------------------------------

We evaluate the utilization of the GPU to represent the hardware utilization in model inference. 
GPU utilization is the percentage of time the GPU is actively processing tasks. 
A higher GPU utilization indicates that the GPU is more actively involved in processing tasks, while a lower utilization indicates that the GPU is less busy or idle. 

We evaluate the GPU utilization during inference with DCN, DCNv2, Wide\&Deep, and DeepFM at the largest computational scale. 
As shown in Figure~\ref{fig: GPU Utilization}, the GPU utilization in \ours~is significantly higher than PyTorch. 
It is because PyTorch schedules all small computational load operators into a single CUDA stream, while \ours~performs intra-module and inter-module parallel execution to fully utilize the GPU. 
Compared to PyTorch, \ours~achieves an average GPU utilization improvement of 24.8\% and 33.3\% on the Avazu and Criteo datasets, with maximum improvements of 36.4\% and 60.6\%, respectively. 
The performance improvement is more pronounced on the Criteo dataset, indicating that \ours~performs better on large data scenarios. 
DCNv2 has the highest GPU utilization during inference among the four models, as it has the largest matrix multiplication workload.

%-------------------------------------------------------------------------------

\subsection{Embedding Sensitivity Studies}
\label{sec: Embedding Sensitivity Studies}
%-----------------------------------

Embedding is an indispensable component in almost all CTR models, and optimization strategies for it remain a hot topic. 
Therefore, we conducted a detailed evaluation of \ours's speedup effect on embedding in several complex scenarios. 
In addition to PyTorch, we also chose TorchRec and OneFlow as baselines, as they have implemented outstanding optimizations for embedding. 

We define the effect of different change factors on embedding as embedding sensitivity.
The inference performance of embedding lookup is mainly related to four factors: 
(1) Batch Size, which represents the number of samples to be inferred; a larger batch size results in a higher output matrix height. 
(2) Embedding Dimension, which is the dimension of the embedding table; a larger dimension results in a wider embedding table and, thus, a wider output matrix. 
(3) Number of Feature Fields, which affects the number of embedding tables and the width of the output matrix. 
(4) Number of Features, which represents the number of features in a feature field; a larger number increases the height of the embedding table, but does not affect the size of the output matrix. 
The experimental results are shown in Figure~\ref{fig: emb_sensitivity}.

\textbf{Batch Size.} 
We set embedding dimension to 32, and the experiments are performed on the Criteo dataset. 
Results show that for different batch sizes, \ours~achieves average speedups of 12.1$\times$, 6.5$\times$, and 7.9$\times$, and maximum speedups of 23.0$\times$, 11.2$\times$, and 13.0$\times$ compared to PyTorch, TorchRec, and OneFlow, respectively. 
It is worth noting that when the batch size reaches 16,384 and above, the performance of TorchRec and OneFlow actually falls behind PyTorch, and their inference latency shows a clear increasing trend. 

\textbf{Embedding Dimension.} 
We set batch size to 2048, and experiments are performed on the Criteo dataset. 
Under different embedding dimensions, \ours~achieves average speedups of 12.9$\times$, 6.2$\times$, and 8.1$\times$ and maximum speedups of 23.0$\times$, 11.3$\times$, and 13.7$\times$ compared to PyTorch, TorchRec, and OneFlow, respectively. 
Results show that the impact of the embedding dimension on the inference latency is relatively small. 
Therefore, in the overall CTR model, as computational scale increases, the proportion of inference latency contributed by the neural network will increase, leading to acceleration degradation and be consistent with the description in~\ref{sec: Speedup}.

\begin{figure}
\centering
  \includegraphics[width=0.995\linewidth]{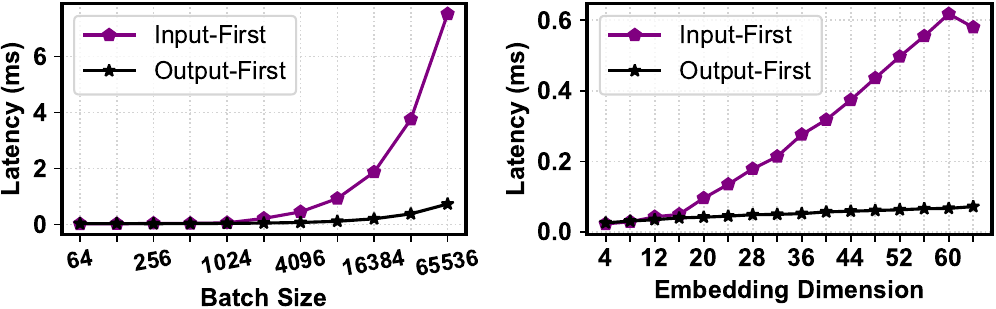}
  \caption{The embedding performance comparison.}
  \label{fig: emb_workload}
\end{figure}

\textbf{Number of Feature Fields.} 
We set batch size and embedding dimension to 2048 and 32, respectively, and the number of features per feature field to 500,000. 
For different numbers of feature fields, \ours~achieves average speedups of 12.2$\times$, 5.6$\times$, and 7.5$\times$, and maximum speedups of 15.4$\times$, 11.0$\times$, and 16.7$\times$, compared to PyTorch, TorchRec, and OneFlow, respectively. 
Results show that as the number of feature fields increases, TorchRec, OneFlow, and \ours~can all maintain stable inference performance. 
However, when the number of feature fields is less than 20, the performance of TorchRec and OneFlow is even worse than PyTorch.

\textbf{Number of Features.} 
We set batch size and embedding dimension to 2048 and 32, respectively, and the number of feature fields to 50. 
For different numbers of features, \ours~achieves average speedups of 12.5$\times$, 5.1$\times$, and 6.5$\times$, and maximum speedups of 13.0$\times$, 5.2$\times$, and 6.6$\times$ compared to PyTorch, TorchRec, and OneFlow, respectively. 
Results show that the number of features has almost no effect on the inference latency of embedding since it is not related to the size of the input or output.

\textbf{Workload Allocation.} 
Finally, we compared two different workload allocation strategies mentioned in~\ref{sec: Workload Allocation}. 
We perform experiments on the Criteo dataset. 
Figure~\ref{fig: emb_workload} shows that when the workload is small, the two allocation strategies have similar inference performance. 
However, as the workload increases, the shortcomings of the input-first allocation mode become apparent. 
In this mode, the memory addresses for reading and writing data between threads are not contiguous. 
In contrast, our output-first allocation can maintain stable performance, especially when the batch size is 65,536 and when the embedding dimension is 60, the inference performance difference between the two strategies is 10.3$\times$ and 9.2$\times$.

%-------------------------------------------------------------------------------

\subsection{Evaluation for Stream Scheduling Strategy}
\label{sec: Evaluation for Scheduling Strategy}
%-----------------------------------

To validate the effectiveness of the breadth-first multi-stream scheduling, we compare it with single-stream scheduling and normal multi-stream scheduling. 
The evaluation metric is the latency from the launch of the first operator in neural network to the point when both parallel branches begin execution. 
For example, when the first operators of explicit interaction module and implicit interaction module are \textit{OP\_Explicit\_0} and \textit{OP\_Implicit\_0}, respectively, and scheduling begins with explicit interaction module, the measured result is the latency from the launch of \textit{OP\_Explicit\_0} to the start of execution for \textit{OP\_Implicit\_0}. 

\begin{figure}
\centering
  \includegraphics[width=0.75\linewidth]{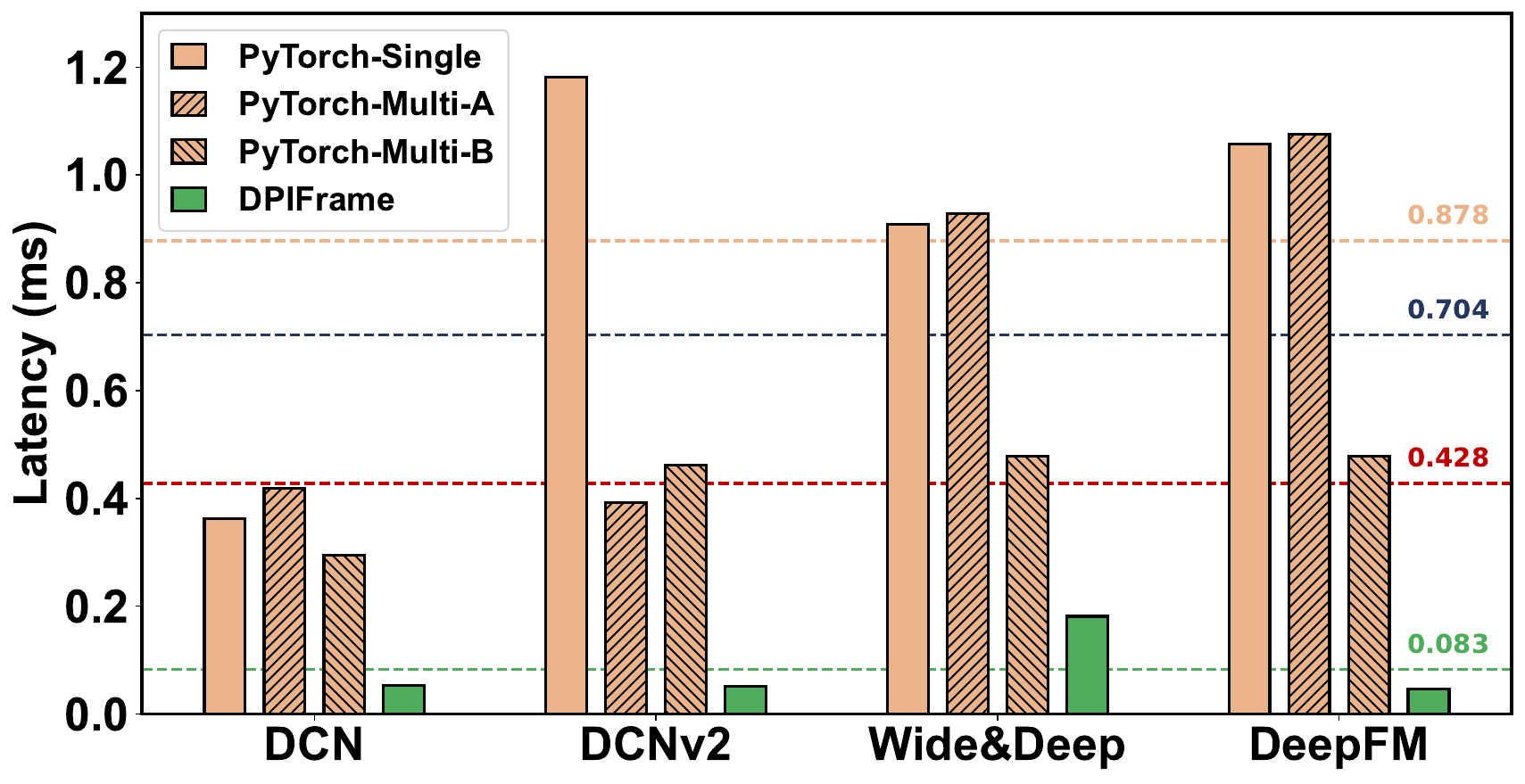}
  \caption{The latency from the launch of the first operator in neural network to the point where both parallel branches begin execution. 
  PyTorch-Single represents the default single-stream scheduling, while PyTorch-Multi-A and PyTorch-Multi-B represent the multi-stream scheduling of PyTorch. 
  The former starts scheduling with explicit interaction module, while the latter begins with implicit interaction module.}
  \label{fig: stream_latency}
\end{figure}

The evaluation is conducted at the largest computational scale on the Criteo dataset as shown in Figure~\ref{fig: stream_latency}. 
Experimental results show that \ours's breadth-first multi-stream scheduling significantly outperforms other approaches, with a potential performance difference of up to 23.18$\times$ compared to the default single-stream scheduling. 
The significant improvement benefits from both parallel branches of neural network begin executing as early as possible. 
This increases opportunities for parallel computation and reduces overall inference latency.

%-------------------------------------------------------------------------------

\subsection{Startup Sequence about Configuration}
\label{sec: Discussion}
%-----------------------------------

\begin{figure}
\centering
  \includegraphics[width=0.75\linewidth]{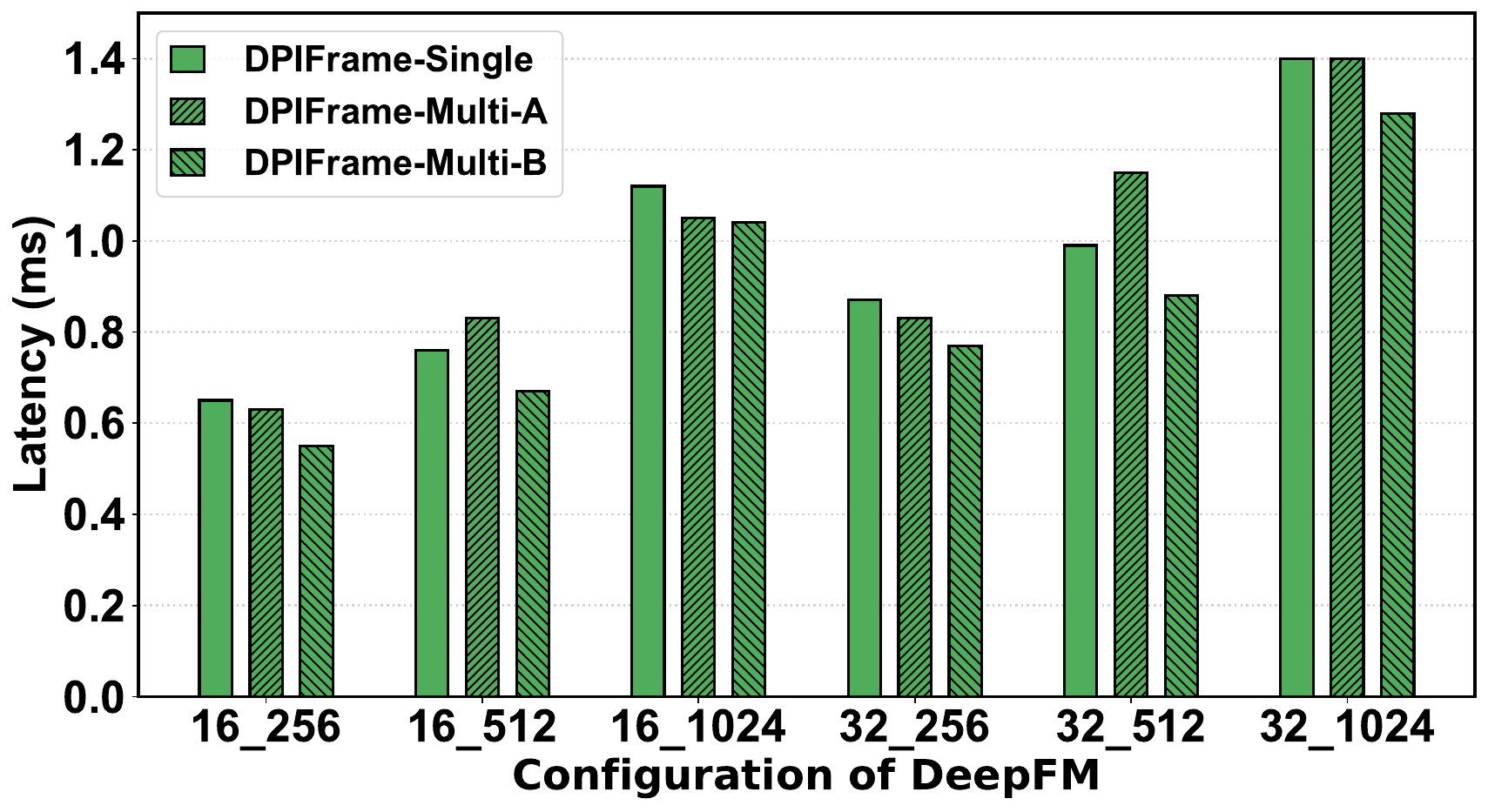}
  \caption{The inference latency of DeepFM with different scheduling strategies. 
  \ours-Single represents the single-stream scheduling, while \ours-Multi-A and \ours-Multi-B use breadth-first multi-stream scheduling. 
  The former starts scheduling with implicit interaction module, while the latter begins with explicit interaction module.
  }
  
  \label{fig: discussion}
\end{figure}

We observe a phenomenon worth pointing out when the start order of the multi-stream scheduling branches is changed. 
As shown in Figure~\ref{fig: discussion}, \ours~generally achieves significant performance improvements. 
However, for DeepFM with an MLP hidden size of 512, \ours-Multi-A shows a worrisome negative optimization result. 
Using the NVIDIA Nsight Systems profiling tool, we find that the cause of this problem is the behaviour of cuBLAS, which is used to implement GEMM. 

To be more specific, when using the same cuBLAS API, changes in the matrix shape can lead to different underlying CUDA kernel selections. 
These differences include not only the kernel name, but also configuration parameters such as grid size, block size, and sometimes even additional \textit{Memset} operations to pad or reorganize data to match the expected matrix formats. 
This problem occurs when DeepFM is configured with 16\_512 or 32\_512. 
a) The GEMM-related CUDA kernel requires additional CUDA operations like \textit{Memset}, and when implicit interaction module is started first, this additional latency cannot be overlapped; 
b) The large grid size of the kernel, consisting of hundreds of blocks, means that the GPU's Streaming Multiprocessors (SMs) are heavily utilized by this kernel. 
As a result, when the inner product is launched later, it will find SMs with limited availability, resulting in significantly higher latency. 
Even setting different priorities for the two CUDA streams does not solve the problem. 
As NVIDIA CUDA officially clarifies, pending work in higher priority streams precedes pending work in lower priority streams~\cite{guide2020cuda}. 
However, it does not guarantee that a kernel from higher priority CUDA streams can preempt SMs already assigned to a running kernel.

\ours-Multi-B starts with explicit interaction module, allowing implicit interaction module to execute as early as possible without compromising the performance of the inner product operator. 
It can also overlap the latency of additional CUDA operations such as \textit{Memset}. 
This results in a consistent acceleration effect. 
Therefore, it is important to consider multiple factors when implementing multi-stream scheduling, including differences in kernel-related parameters and the actual scheduling behaviour of the hardware.

\section{RELATED WORK}
\label{sec: Related Work}
%-------------------------------------------------------------------------------

\textbf{CTR Prediction Models.} 
Recently, with the rapid advancement of deep learning technologies, deep neural network models with embedding modules have been applied to CTR prediction, yielding significant improvements in accuracy. 
For example, Wide\&Deep~\cite{cheng2016wide} first uses embedding layers to learn low-dimensional dense representations from sparse input and then designs task-specific fully connected layers to capture the latent interactions between features. 
DeepFM~\cite{guo2017deepfm} uses factorization machines to model both low and high-order feature correlations. 
DCN~\cite{wang2017deep} and DCNv2~\cite{wang2021dcn} leverage cross network to perform explicit interaction effectively. 
Overall, these models adopt the paradigm of embedding and neural network, significantly reducing the need for feature engineering and increasing the expressiveness of the models. 

\textbf{Optimizations for Embedding.} 
Hashing methods~\cite{desai2022random, pansare2022learning} allow multiple elements to be mapped to the same embedding vector, thus reducing the embedding space. 
However, hashing conflicts can lead to a significant loss of accuracy. 
ALPT~\cite{li2023adaptive} develops an adaptive low-precision training method to successfully train 8-bit embeddings without sacrificing the prediction accuracy of the CTR model. 
DeepLight~\cite{deng2021deeplight} prunes the dense embedding vectors to make them sparse in the embedding matrix. 
These methods all modify the original parameters or the computation process of the embedding, which inevitably affects the accuracy. 
Therefore, \ours~introduces a multi-table parallel embedding lookup algorithm that achieves significant acceleration without modifying the embedding itself.

\textbf{GPU Acceleration for DLRMs.} 
EL-Rec~\cite{wang2022rec} uses tensor-train decomposition to compress embedding tables, enabling data-parallel training. 
RECom~\cite{pan2023recom} is a compiler approach that fuses massive embedding columns into a single GPU kernel. 
FEC~\cite{ma2023fec} uses embedding tiering with AllReduce for hot entries and prefetching for cold entries. 
OPER~\cite{wang2024oper} proposes optimality-guided embedding table parallelization to balance workload and communication.
RecFlex~\cite{pan2024recflex} introduces feature heterogeneity-aware optimization, generating fused kernels with distinct schedules for different feature fields. 
To address input preprocessing overhead, RAP~\cite{wang2024rap} enables resource-aware GPU sharing for preprocessing and training overlap.
NDRec~\cite{li2024ndrec} offloads embedding operations to computational storage devices with lookahead embedding.
TRACI~\cite{huang2025traci} proposes in-network acceleration for the aggregation operator, leveraging input and output reuse. 
Our \ours~based on a single GPU is the first to identify the critical limitation of CTR model inference inefficiency as the mismatch between serial computational pattern and parallel model structure. 
Then innovatively propose a dual parallelizable approach that performs parallel computation both intra-module and inter-module. 

%-------------------------------------------------------------------------------

\section{CONCLUSION}
\label{sec: Conclusion}
%-------------------------------------------------------------------------------

In this paper, we propose \ours, the first dual parallel inference framework for the CTR model that supports both intra-module and inter-module parallel execution to enhance efficiency in CTR prediction. 
\ours~leverages its efficient multi-table parallel lookup algorithm and thoughtful workload allocation to fully utilize hardware capabilities and significantly reduce inference latency. 
The designed breadth-first stream scheduling strategy further addresses the mismatch between hardware execution and model structure by scheduling operators with multiple CUDA streams. 
The experimental results show that our \ours~comprehensively outperforms the comparative baselines. 
Compared with PyTorch, the embedding latency of our \ours~can be reduced by an average of 12.9$\times$ and a maximum of 23.0$\times$.
Compared with PyTorch, TorchRec, HugeCTR, and OneFlow, \ours~can improve the overall performance of CTR model inference by 5.83$\times$, 4.29$\times$, 2.15$\times$, and 2.0$\times$, respectively.
%-------------------------------------------------------------------------------

\renewcommand\bibfont{\footnotesize}
\printbibliography

\vspace{-35pt}
    
\begin{IEEEbiography}[{\includegraphics[height=1.25in,trim=0cm 1cm 0cm 0cm,clip,keepaspectratio]{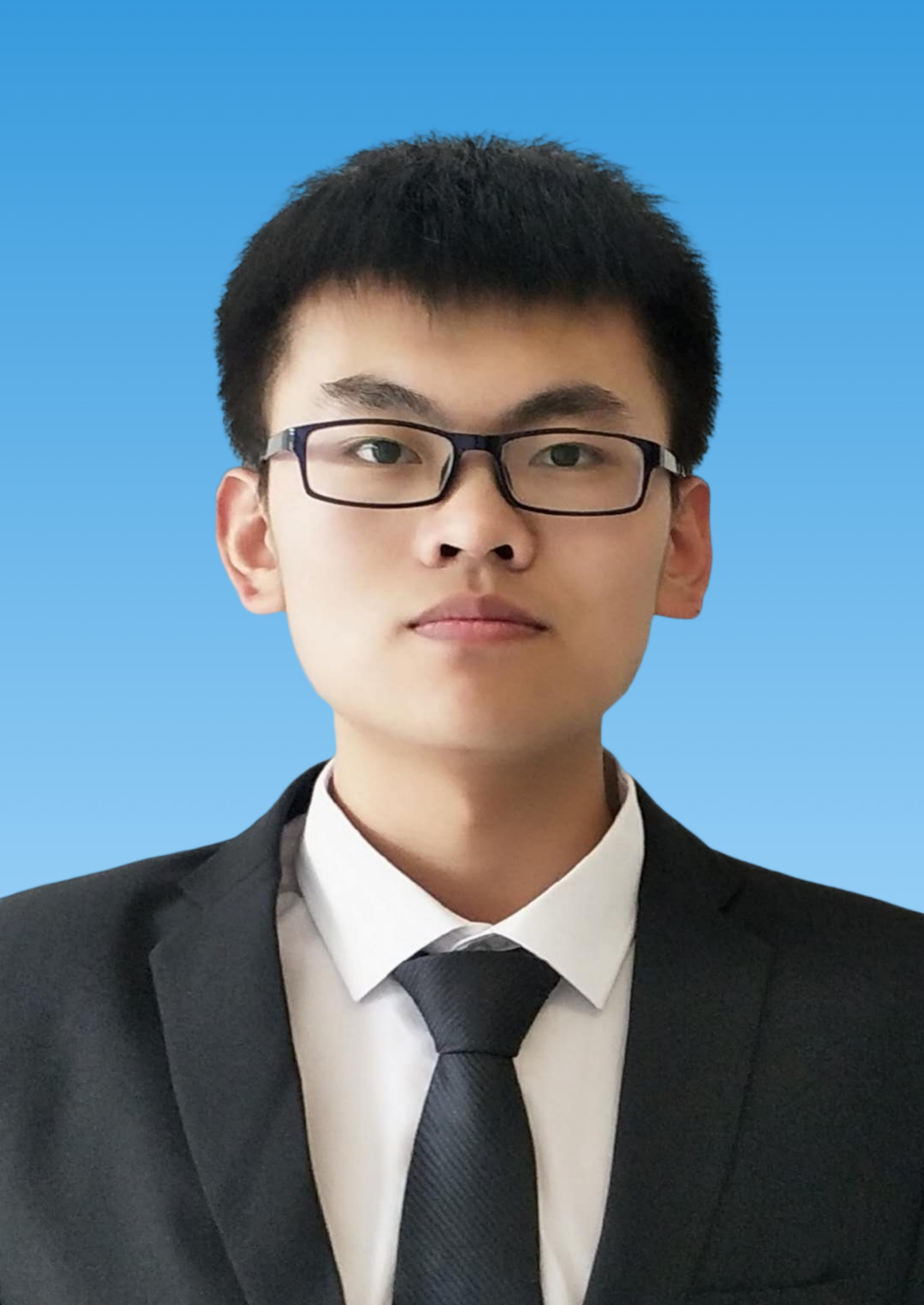}}]{Dezhi Yi}
received the B.Eng. degree in Internet of Things Engineering from Nankai University, Tianjin, China, in 2021. He is currently working toward the Ph.D. degree with the College of Computer Science, Nankai University, Tianjin, China, supervised by Prof. Ye Lu. His research interests include software–hardware codesign and recommender systems.
\end{IEEEbiography}

\vspace{-35pt}

\begin{IEEEbiography}[{\includegraphics[height=1.25in,trim=0cm 1cm 0cm 0cm,clip,keepaspectratio]{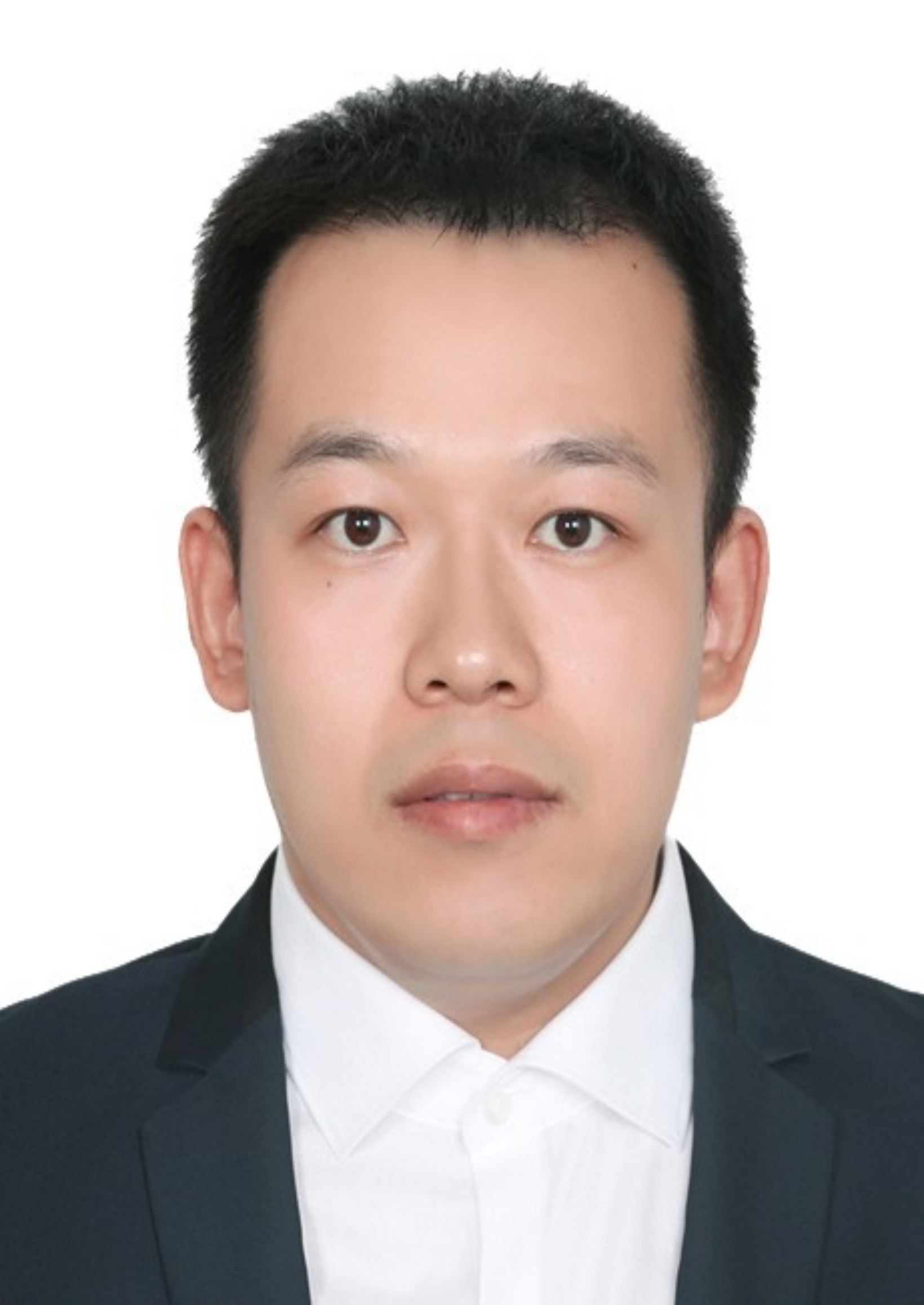}}]{Huifeng Guo}
received his B.S. degree from Lanzhou University in 2012, M.S. and Ph.D. degrees from Harbin Institute of Technology in 2014 and 2018, respectively. He is currently a researcher at Huawei Technologies Co., Ltd., focusing on recommendation, information retrieval, and advertising algorithms and applications based on AI-related technologies.
\end{IEEEbiography}

\vspace{-35pt}

\begin{IEEEbiography}[{\includegraphics[height=1.25in,trim=0cm 1cm 0cm 0cm,clip,keepaspectratio]{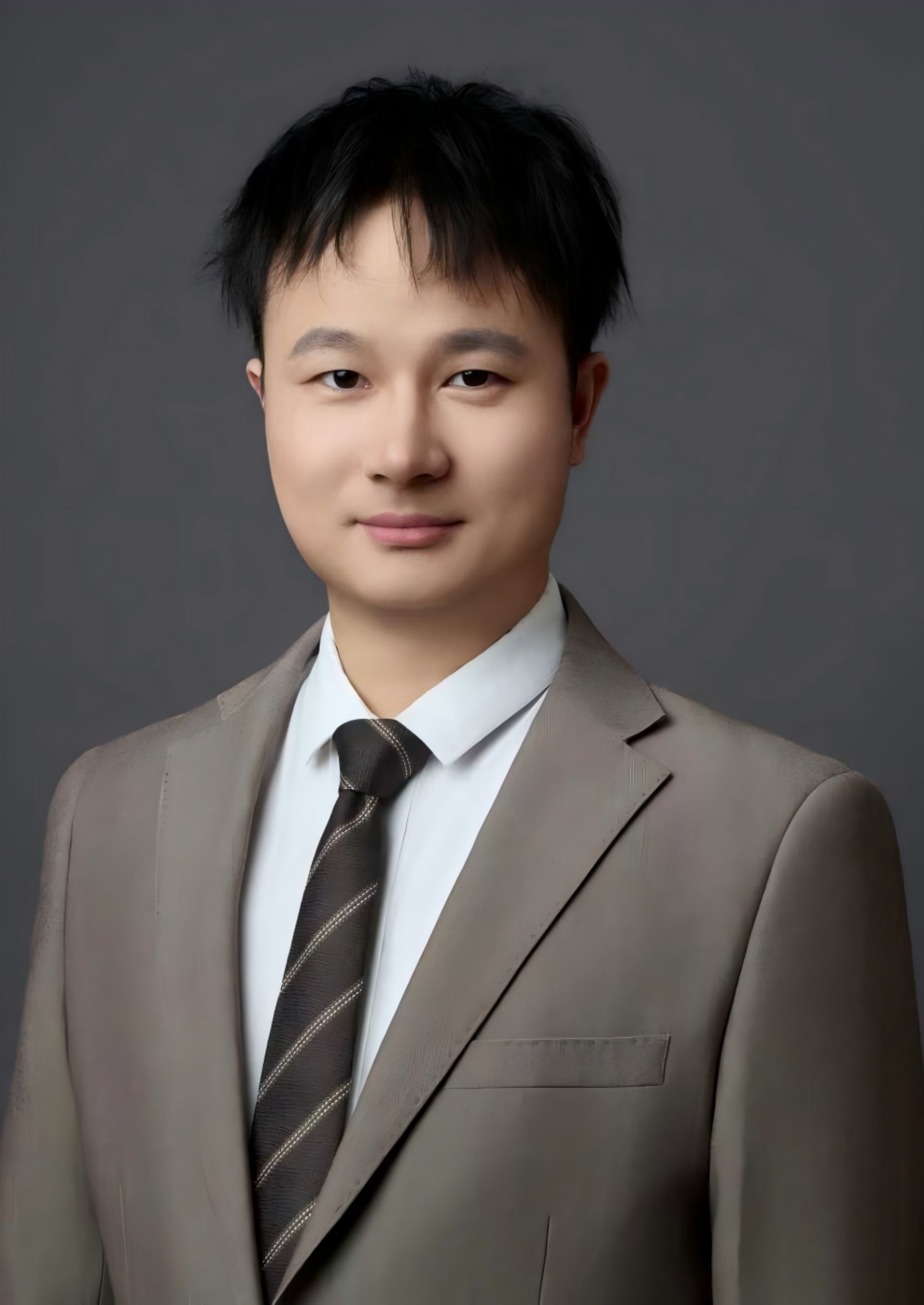}}]{Kunpeng Xie}
received the B.S. and Ph.D. degrees from Nankai University, Tianjin, China, in 2019 and 2024, respectively. His research interests include computer architecture and hardware design.
\end{IEEEbiography}

\vspace{-35pt}

\begin{IEEEbiography}[{\includegraphics[height=1.25in,trim=0cm 1cm 0cm 0cm,clip,keepaspectratio]{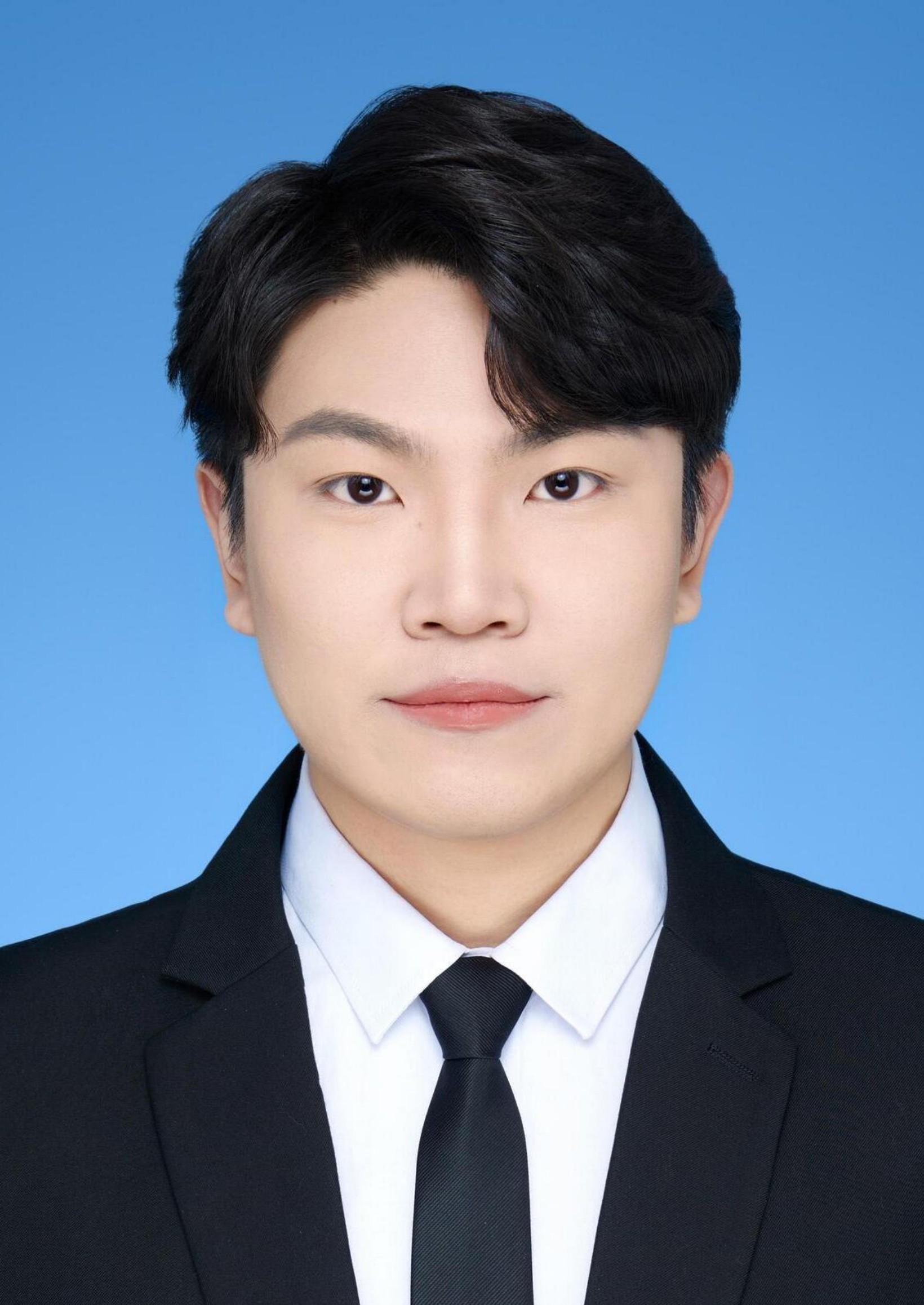}}]{Zhaolong Jian}
received the B.Eng. degree in Internet of Things Engineering from Nankai University, Tianjin, China, in 2020. He is currently working toward the Ph.D. degree with the College of Computer Science, Nankai University, Tianjin, China. His main research interests include computer system architecture, trusted execution environment, and system security.
\end{IEEEbiography}

\vspace{-35pt}

\begin{IEEEbiography}[{\includegraphics[height=1.25in,trim=0cm 1cm 0cm 0cm,clip,keepaspectratio]{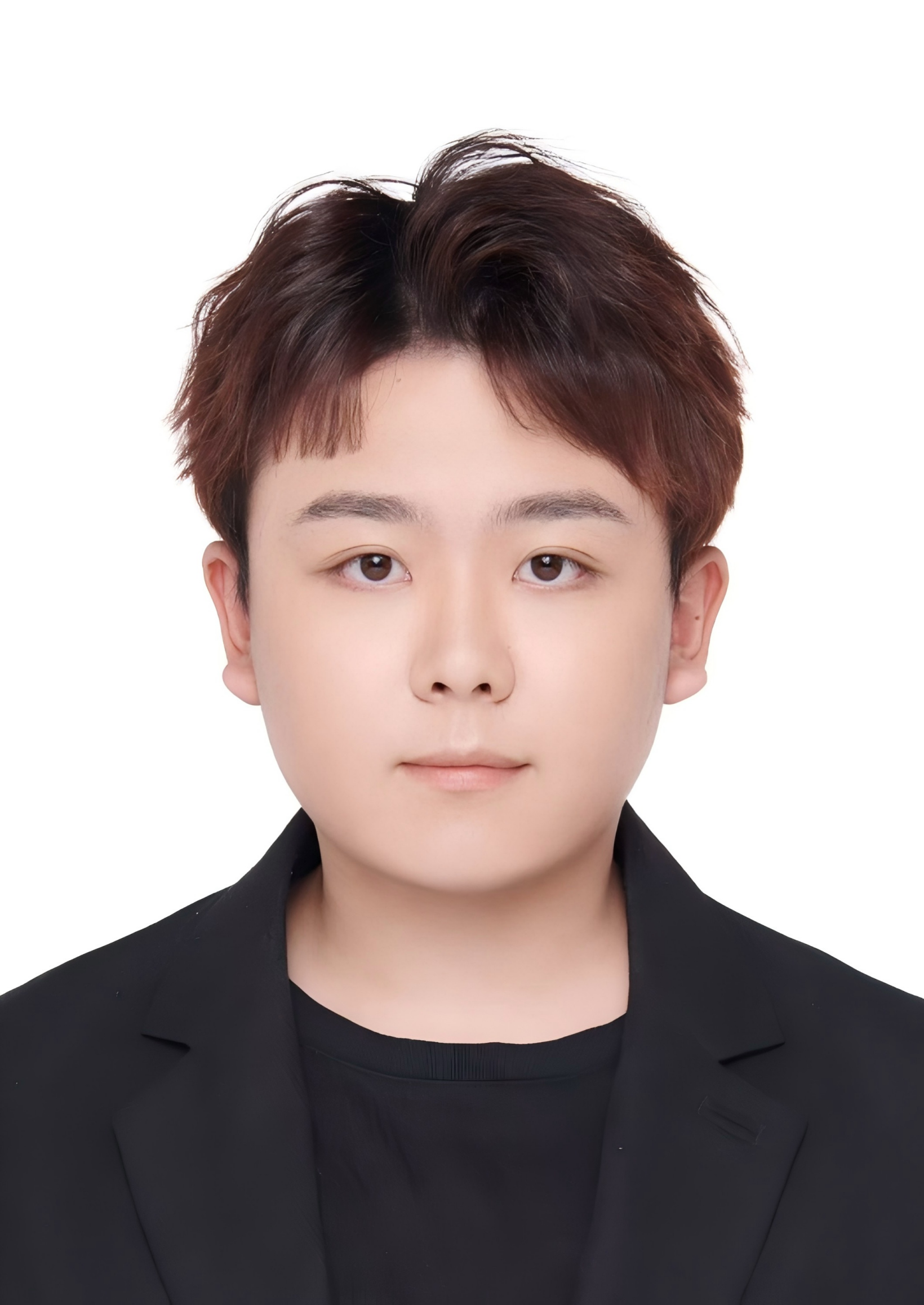}}]{Haochi Yu}
received the B.Eng. degree in Information Security from Nankai University, Tianjin, China, in 2024. He is currently working toward the M.S. degree with the College of Computer Science, Nankai University, Tianjin, China, supervised by Prof. Ye Lu. His research interests include inference acceleration and model compression for deep learning systems.
\end{IEEEbiography}

\vspace{-35pt}

\begin{IEEEbiography}[{\includegraphics[height=1.25in,trim=0cm 1cm 0cm 0cm,clip,keepaspectratio]{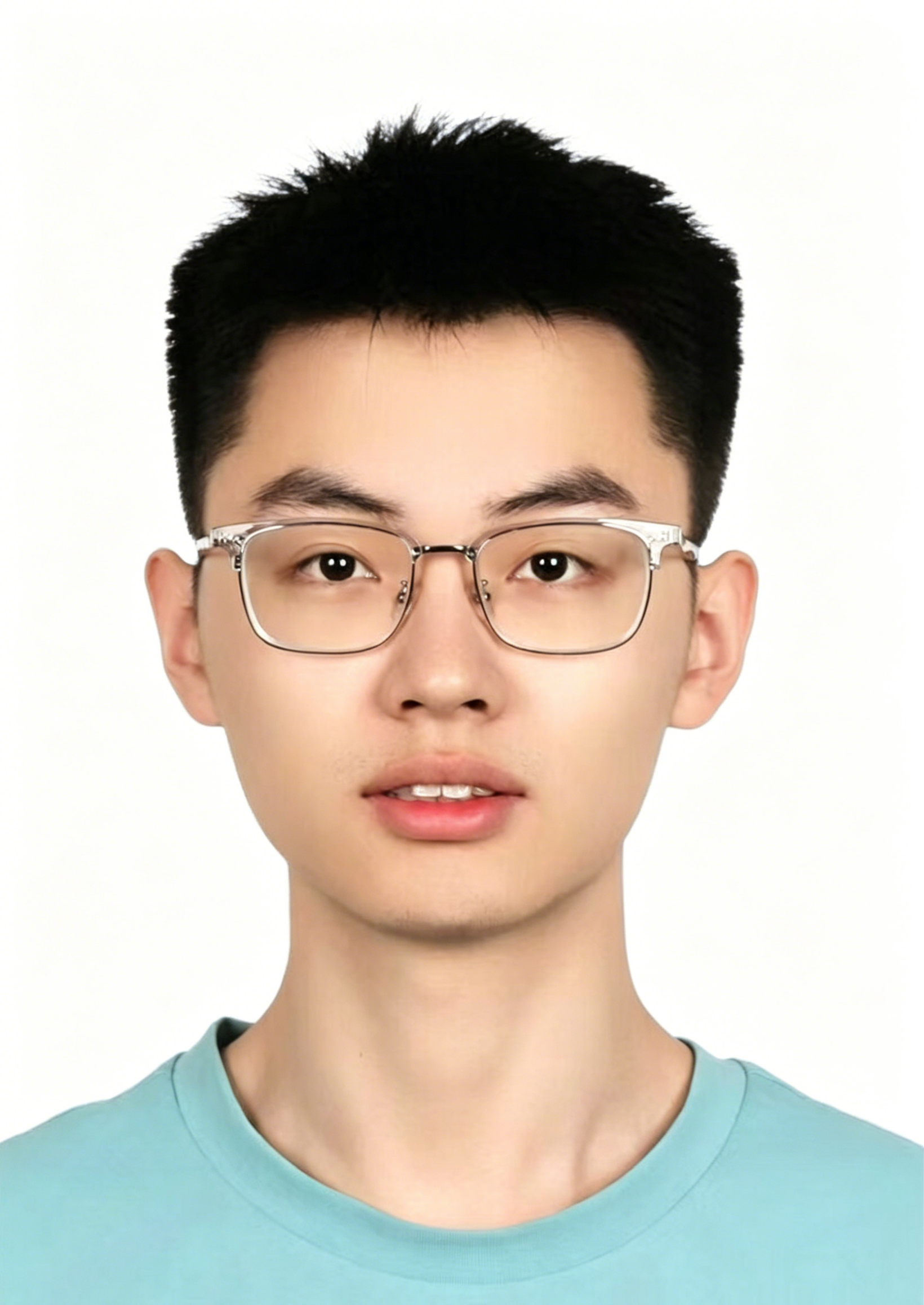}}]{Wenxuan He}
received the B.Eng. degree in Information Security from Nankai University, Tianjin, China, in 2024. He is currently working toward the M.S. degree with the College of Computer Science, Nankai University, Tianjin, China, supervised by Prof. Ye Lu. His research interests include optimization algorithms and system architecture.
\end{IEEEbiography}

\vspace{-35pt}

\begin{IEEEbiography}[{\includegraphics[height=1.25in,trim=0cm 1cm 0cm 0cm,clip,keepaspectratio]{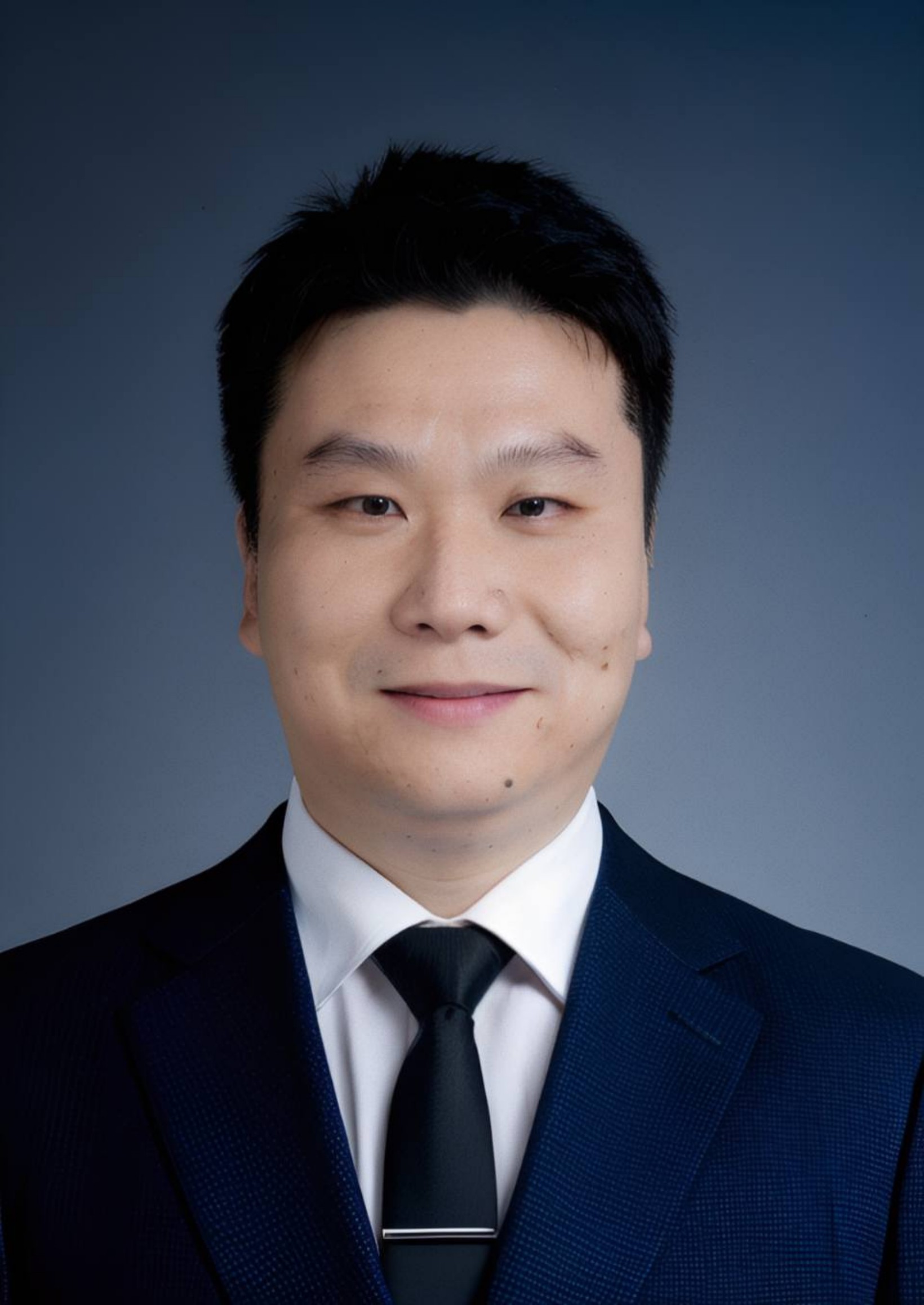}}]{Zhenhua Dong}
received Ph.D. degrees from Nankai University, Tianjin, China, in 2012. His research interests include Recommender system, information retrieval, causal inference, AI infrastructure.
\end{IEEEbiography}

\vspace{-35pt}

\begin{IEEEbiography}[{\includegraphics[height=1.25in,trim=0cm 1cm 0cm 0cm,clip,keepaspectratio]{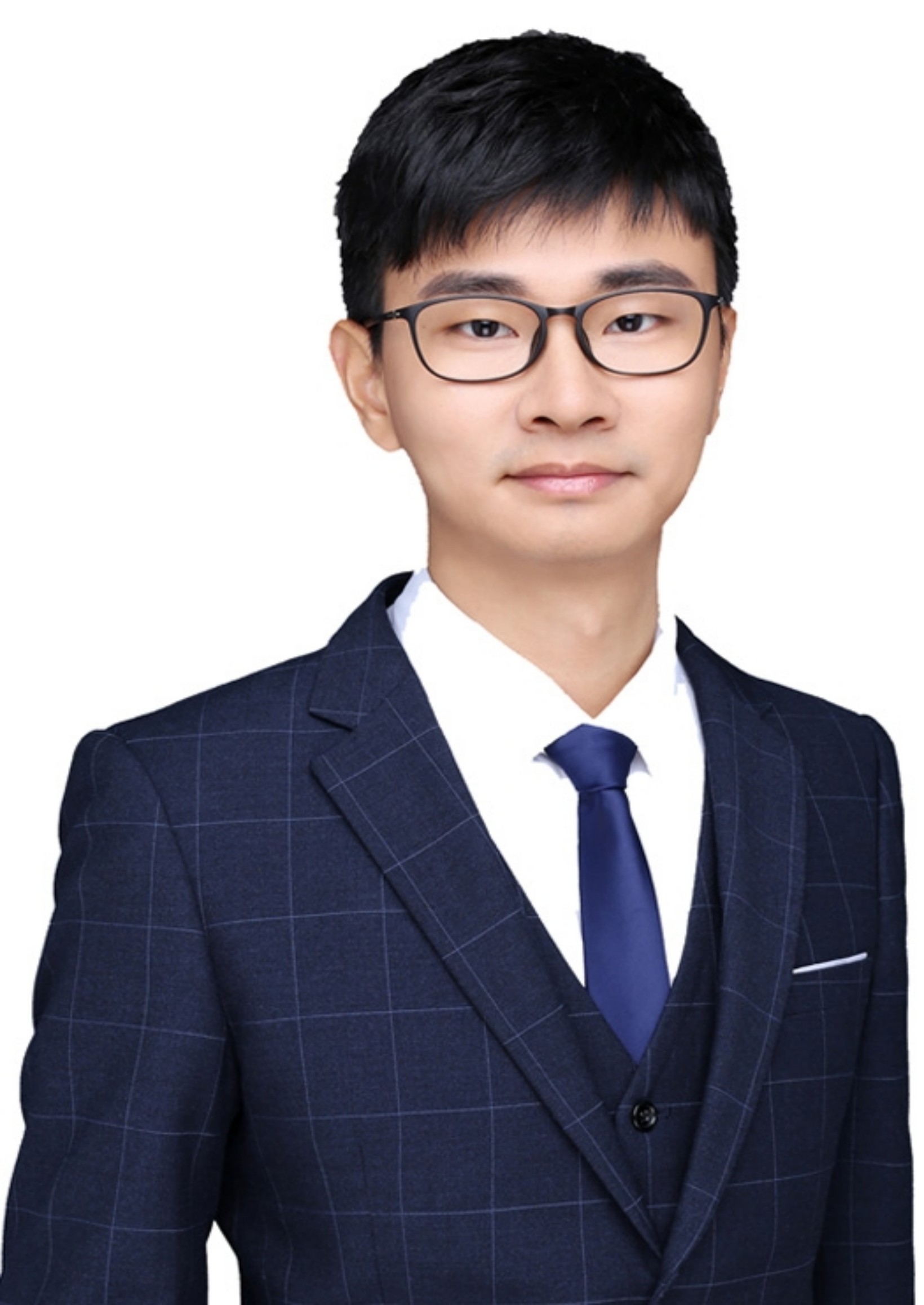}}]{Ruiming Tang}
is a technical expert in recommendation area and is assigned as the director of ranking model center and large language model center in Kuaishou. He was working in Huawei Noah’s Ark Lab from Dec 2014 to Jun 2025 and was assigned as the director of recommendation and search lab. He got his Ph.D. degree from National University of Singapore in 2014. His research directions include (but not limited to) generative recommendation, user behavior modelling, LLM enhanced recommendation, long-context in LLM, etc.
\end{IEEEbiography}

\vspace{-35pt}

\begin{IEEEbiography}[{\includegraphics[height=1.25in,trim=0cm 1cm 0cm 0cm,clip,keepaspectratio]{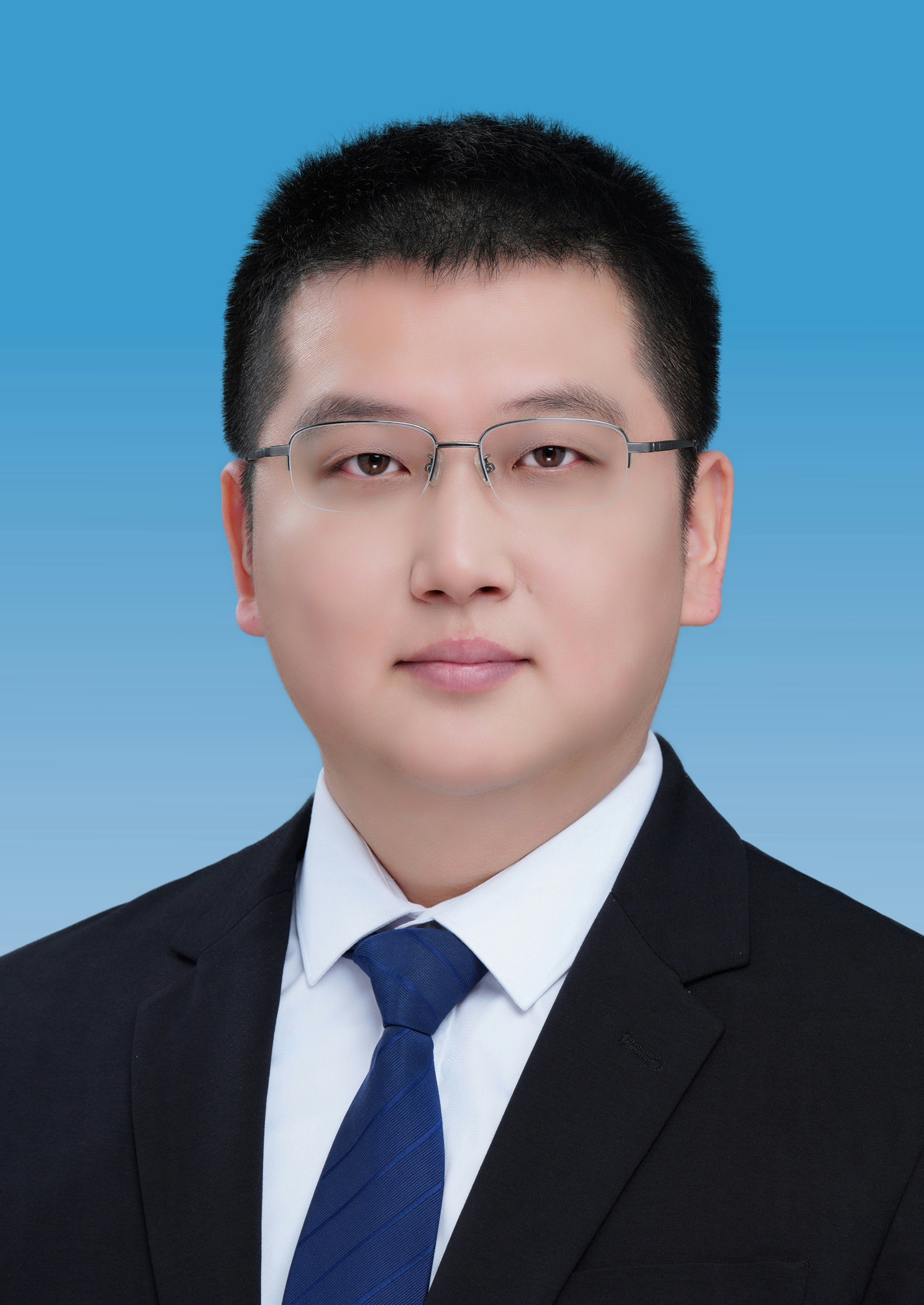}}]{Ye Lu}
received the B.S. and Ph.D. degrees from Nankai University, Tianjin, China, in 2010 and 2015, respectively. Currently, he is a Professor with the College of Cryptology and Cyber Science, Nankai University. His research interests include computer architecture, artiﬁcial intelligence system, and system security.
\end{IEEEbiography}

\end{document}